\documentclass{npcs}
\usepackage{axodraw}
\usepackage{graphics} 
\usepackage{epsfig}
\usepackage{multicol}
\usepackage{amsmath}
\newcommand{\multsp}{\,}

\begin{document}

\runauthor{T. Shishkina, V. Mossolov, I. Marfin} \runtitle{The investigation of the $\gamma\gamma\rightarrow 
W^+W^-$ including electromagnetic corrections 
at the TESLA kinematics}
\begin{topmatter}
\title{ The investigation of the $\gamma\gamma\rightarrow
W^+W^-$ including electromagnetic corrections 
at the TESLA kinematics}
\author{Author I.B. Marfin}
\institution{NCHEP}
\address{153 Bogdanovitcha str.,220040 Minsk, Belarus}
\email{marfin@hep.by}
\author{Author V.A. Mossolov}
\institution{NCHEP}
\address{153 Bogdanovitcha str.,220040 Minsk, Belarus}
\email{mos@hep.by}
\author{Author T.V. Shishkina}
\institution{NCHEP}
\address{153 Bogdanovitcha str.,220040 Minsk, Belarus}
\email{shishkina@hep.by}
\begin{abstract}

The $W^+W^-$ production in $\gamma\gamma$ scattering is considered in the Standard Model.
The main contribution to radiative effects for the  process $\gamma\gamma\rightarrow W^+W^-$ has been
calculating and analyzing. It is found the latter is  considerable at
high energies and greatly contributes to the differential cross
section $d\sigma_{\lambda_1\lambda_2\lambda_3\lambda_4}$ at various
polarizations of initial photons and final bosons. Monte-Carlo generator built based on TESLA kinematics. 
\end{abstract}

\end{topmatter}
\newpage

\section{Introduction}

$ $

The multiple vector-boson production will be a
crucial test of gauge structure of the Standard Model since the
triple and quartic vector-boson couplings involved in this kind of
reaction are constrained by the $SU(2)\bigotimes U(1)$ gauge
invariance. The production of several vector bosons is the best place to
search directly for any anomalous behavior of triple and quartic
couplings \cite{bib0}.

Any small deviation from the Standard Model predictions for these
coupling spoils the cancellations of the high energy
behaviour between the various diagrams, giving rise to an
anomalous growth of the cross section with energy. It is important
to measure the vector-boson selfcouplings and look for deviations
from the Standard Model.

Using the linear $e^+e^-$-collider, one can
obtain the colliding $\gamma e$ and $\gamma\gamma$ beams with
 the same energies as in $e^+e^-$-
collisions and with high luminosity.

Thus, there is the possibility of direct experimental
investigation of the reactions: $$\gamma e\rightarrow W\nu $$ and
$$\gamma\gamma \rightarrow W^+W^-. $$ These reactions were
considered as parts of the W-bosons production process in $e^+e^-$
collisions ($ee\rightarrow eW\nu, e^+e^-\rightarrow W+X,
ee\rightarrow eeWW$) and in $ep$ and $pp$ collisions as well.
However, the corresponding cross sections are very small and such
processes will  be observed in the nearest future hardly.

Reactions $e^+e^-\rightarrow W^+W^-$ and $\gamma\gamma\rightarrow
W^+W^-$ will give the unique possibilities of $ W$-investigation due to
relatively large cross sections and low background. Moreover, the
colliding $\gamma e$ and $\gamma\gamma$ beams will allow us to
study a number of problems which are harding solved in other
collisions \cite{bib3}. The cross sections of process
$\gamma\gamma\rightarrow W^+W^-$ is of order $\sim 4\pi\alpha^2/M_W^2$
 at high energy (they are constant at $s\rightarrow \infty$
due to the t-channel vector exchange). Therefore, at $s_{\gamma
e}\sim s_{\gamma\gamma} \sim s_{ee}\gg M_W^2$, the cross sections
of this reaction considerably exceed the cross section of the
$e^+e^-$ annihilation into hadrons.

\section{The probing triple and quartic gauge bosons couplings}

$ $

The cross section $\gamma\gamma\to W^{+}W^{-}$ is about
an order of magnitude larger than $e^+ e^- \to W^{+}W^{-}$.

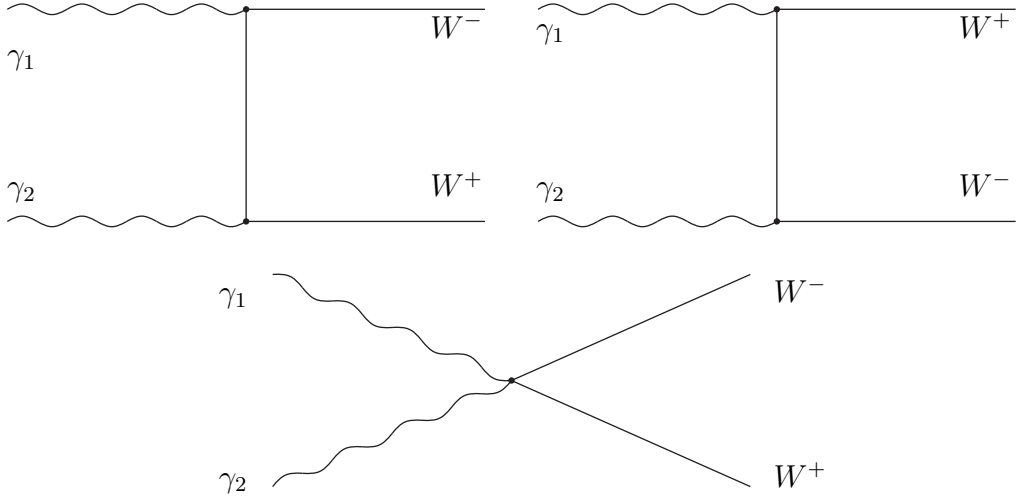
\begin{figure}[h]
\begin{center}
\begin{picture}(400,200)(0,0)

\Photon(10,110)(100,110){2}{4}\put(10,120){$\gamma_2$}
\Vertex(100,110){1.2}
\Line(100,110)(190,110)\put(170,120){$W^+$}
\Line(100,110)(100,190)
\Photon(10,190)(100,190){2}{4}\put(10,170){$\gamma_1$}
\Vertex(100,190){1.2}
\Line(100,190)(190,190)\put(170,180){$W^-$}

\Photon(210,110)(300,110){2}{4}\put(210,120){$\gamma_2$}
\Vertex(300,110){1.2}
\Line(300,110)(390,110)\put(370,120){$W^-$}
\Line(300,110)(300,190)
\Photon(210,190)(300,190){2}{4}\put(210,180){$\gamma_1$}
\Vertex(300,190){1.2}
\Line(300,190)(390,190)\put(370,180){$W^+$}

\Photon(110,10)(200,50){2}{4}\put(90,10){$\gamma_2$}
\Vertex(200,50){1.2}
\Photon(110,90)(200,50){2}{4}\put(90,80){$\gamma_1$}
\Line(200,50)(290,10)\put(300,10){$W^+$}
\Line(200,50)(290,90)\put(300,80){$W^-$}

\end{picture}
\end{center}
\caption{The Feynman diagrams for $W^+W^-$-production}\label{p1}
\end{figure}

We discuss how one can use $\gamma\gamma$ collisions to study
the pair production of vector bosons (see fig. \ref{p1}) $$\gamma+\gamma\ \rightarrow W^{+} + W^{-}$$
and get information on three- and four-boson
vertices. 

The process $\gamma\gamma\rightarrow W^+W^-$ depends at the tree level both
on the triple $\gamma WW$ and the quartic $\gamma\gamma WW$ coupling,
and no other vertices are involved in the unitary gauge at the lowest order \cite{bib1}.
The sensitivity to the $\gamma WW$ coupling is comparable and complementary
to the reactions $e^+e^- \to W^+W^-$ and $\gamma e^- \to W^-\nu$:
the first process involves a mixture of the $\gamma WW$ and the $ZWW$ coupling,
the second -- involves the $\gamma WW$ alone but is not such sensitive.
Because the sensitivity to the $\gamma\gamma WW$ coupling is much larger in the considering process
than the one in  $e^+e^-$ collisions, $\gamma\gamma\to W^+W^-$
is the ideal process to study this coupling.

The measurement
of cross section and asymmetries of the process
$\gamma\gamma\rightarrow W^+W^-Z$ is complimentary to the analysis
of the production of vector-boson pairs.


$ $

We have the  unique possibility to examine production of two and three
vector bosons in $\gamma\gamma$ collisions through the reaction
using beams of polarized and unpolarized photons.

\begin{itemize}
\item This process involves only interactions between the gauge
bosons making more evident any deviation from predictions of the
gauge structure by Standard Model.
\item There is no tree-level contribution involving the Higgs
boson with any uncertainties coming from the scalar
sector.
\end{itemize}

We started from explicit expression for the
amplitude of process $\gamma\gamma\rightarrow W^+W^-$ 

\begin{align}\label{c1}
M = & G_v \epsilon_{\mu}(k_1)\epsilon_{\nu}(k_2)
\epsilon_{\alpha}(p_{+})\epsilon_{\beta}(p_{-})
M_T^{\mu\nu\alpha\beta},
\end{align}
where
\begin{align}\label{c2}
M_T^{\mu\nu\alpha\beta} = \sum_{i=1}^{3}
M_i^{\mu\nu\alpha\beta},
\end{align}
$k_1$, $k_2$, $p_{+}$, $p_{-}$ are four-momenta of the $\gamma$, $\gamma$, $W^{+}$, $W^{-}$, respectively 
and $\epsilon_{\mu}(k_1), \epsilon_{\nu}(k_2),\epsilon_{\alpha}(p_{+}), \epsilon_{\beta}(p_{-})$ 
-- their polarizations.
$$G_v = e^3 \cot\theta_W.$$

It is convenient to define the triple-gauge-boson coupling in the following way:
\begin{align}\label{a1}
{\Gamma}_{3}^{\alpha\beta\gamma}(P_1,P_2) = & [(2P_1 +
P_2)^{\beta} g^{\alpha\gamma} - (2P_2 + P_1)^{\alpha}
g^{\beta\gamma} + (P_2 - P_1)^{\gamma} g^{\alpha\beta}],
\end{align}
the quartic-gauge-boson coupling  as
\begin{align}\label{a2}
{\Gamma}_{4}^{\mu\nu\alpha\beta} = & g^{\mu\alpha} g^{\nu\beta} +
g^{\mu\beta} g^{\nu\alpha} -2 g^{\mu\nu} g^{\alpha\beta},
\end{align}
and the propagator tensor for the vector boson     as
\begin{align}\label{a3}
D^{\mu\nu}(k) = & \frac{(g{\mu\nu} - k^{\mu}k^{\nu}/m^2)}{k^2 -
m^2}.
\end{align}
Using the expression defined above, the contributions of the three
Feynman diagrams (see fig. $1$) for the $WW$-production can be written as
\begin{align}\label{b8}
M_{1}^{\mu\nu\alpha\beta} = & {\Gamma}_{3}^{\mu\alpha\xi}
(-k_1,p_{+}) D_{\xi\lambda}(p_+ -
k_1){\Gamma}_{3}^{\nu\beta\lambda}(-k_2,p_-),
\end{align}
\begin{align}\label{b9}
M_{2}^{\mu\nu\alpha\beta} = & {\Gamma}_{3}^{\mu\beta\xi}
(-k_1,p_{-}) D_{\xi\lambda}(p_- -
k_1){\Gamma}_{3}^{\nu\alpha\lambda}(-k_2,p_+),
\end{align}
\begin{align}\label{b10}
M_{3}^{\mu\nu\alpha\beta} = & {\Gamma}_{4}^{\mu\nu\alpha\beta}.
\end{align}
Then we calculate cross sections for polarized beams and for all combinations of
polarizations of final particles. The exact covariant expressions for all amplitudes of different
Feynman diagrams (see fig. \ref{p1}) are presented in the appendix (\ref{ap1}) -- (\ref{ap4}).

\section{Cross sections and asymmetries}

$ $

We analyze the total cross sections of the considered process
as well as different distributions of the final state vector
bosons.  We have evaluated the total cross section for the
processes $\gamma\gamma\rightarrow W^+W^-$
\begin{eqnarray}
\begin{array}{c}
\displaystyle \sigma = \int|M|^2d\Gamma,
\end{array}
\end{eqnarray}
where $|M|$ can be defined through formula (\ref{ap1}) -- (\ref{ap4}),  the phase-space volume element is assumed as
$$d\Gamma=\frac{d^3p_+}{(2\pi)^32p_+^0}\frac{d^3p_-}{(2\pi)^32p_-^0}(2\pi)^4\delta(k_1+k_2-p_+-p_-).$$
The total cross sections  of $\gamma\gamma\rightarrow W^+W^-$ at  $\sqrt{s}\simeq 1$ TeV
is about $\sigma = 3.89\cdot 10^{-8}\ \mbox{GeV}^{-2}$.


In order to reach the best understanding of these reactions we
present here various differential cross-section  as  well as
polarized asymmetries  of two- and three-boson production:
\begin{eqnarray}
\begin{array}{c}
\displaystyle A_1 = \frac{d\sigma_{+,+,+,+}/dy_- - d\sigma_{+,-,+,+}/dy_-}{d\sigma_{+,+,+,+}/dy_- + d\sigma_{+,-,+,+}/dy_-}, \\ 
\displaystyle A_2 = \frac{d\sigma_{+,+,+,-}/dy_- - d\sigma_{+,-,+,-}/dy_-}{d\sigma_{+,+,+,-}/dy_- + d\sigma_{+,-,+,-}/dy_-}, \\
\displaystyle A_3 = \frac{d\sigma_{+,+,0,0}/dy_--d\sigma_{+,-,0,0}/dy_-}{d\sigma_{+,+,0,0}/dy_-+d\sigma_{+,-,0,0}/dy_-}, \\
\displaystyle A_4 = \frac{d\sigma_{+,+,unpol,unpol}/dy_- - d\sigma_{+,-,unpol,unpol}/dy_-}{d\sigma_{+,+,unpol,unpol}/dy_- + d\sigma_{+,-,unpol,unpol}/dy_-}. \\
\end{array}
\end{eqnarray}
Here the following notations are introduced: $+,-$ denote right and left circular polarizations of gammas and final bosons, $0$ means a longitudinal polarization, $unpol$ relates to 
the case of unpolarized particles.

\begin{figure}[h!]
\leavevmode
\begin{minipage}[b]{.475\linewidth}
\centering
\includegraphics[width=\linewidth, height=3.8in, angle=0]{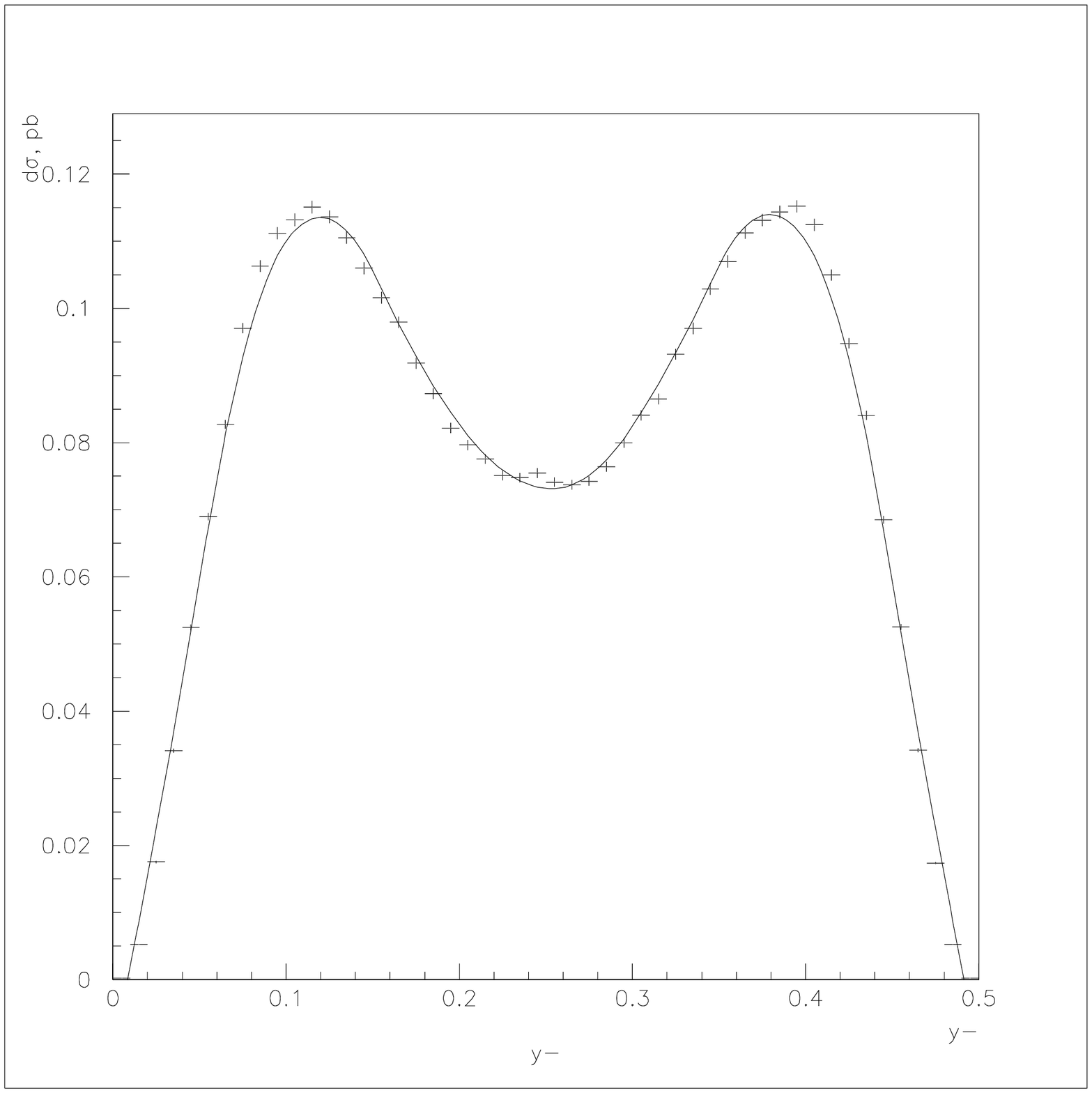}
\caption{ The differential cross section  $d\sigma_{+,+,+,+}/dy_{-}$} \label{p6}
\end{minipage}
\begin{minipage}[b]{.475\linewidth} \centering
\includegraphics[width=\linewidth, height=3.8in, angle=0]{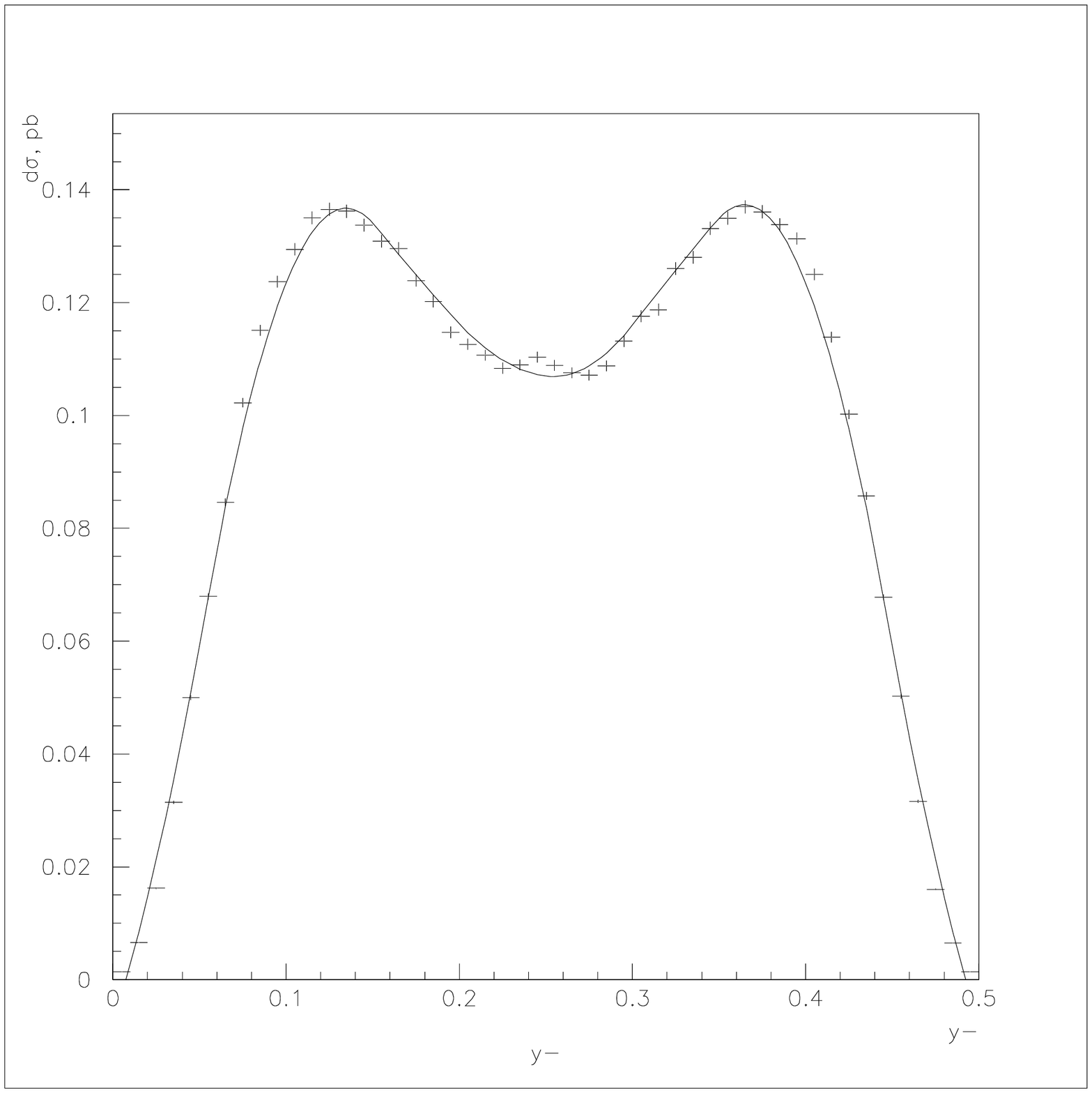}
\caption{ The differential cross section $d\sigma_{+,-,+,+}/dy_{-}$ } \label{p7}
\end{minipage}\hfill
\end{figure}

One can see the $y_-$-distribution($y_-=(k_1,p_-)/S$) and different kinds of
polarized  asymmetries  in fig. \ref{p6} -- fig. \ref{p42}.

Using these  formulas we have  constructed the  Monte-Carlo event`s
generator \cite{bib2}. The universal kinematical block of this
generator allows us to  produce $2$-particles, $3$-particles  and even
more particles final state in    identical  way.  We  have  used
this block to calculation above  mentioned processes  and we  are
going to use it to another similar calculations some others
processes in TESLA kinematics. It is realized as a fast-working
FORTRAN  module which provides required number  of  events (up to
several billions in realistic time) with a  defined initial energy
and final masses.

\begin{center}
\begin{figure}[h!]
 \leavevmode
\begin{minipage}[b]{1.\linewidth}
\centering
\includegraphics[width=0.5\linewidth, height=3.9in, angle=0]{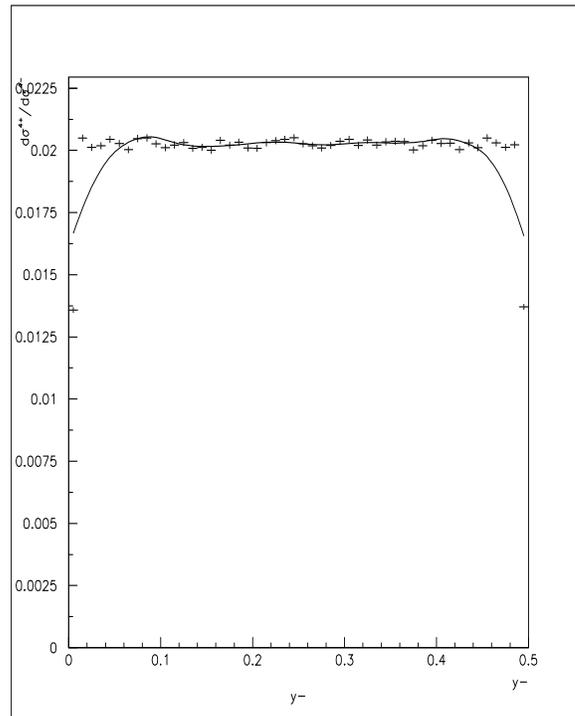}
\begin{center}
\caption{ The ratio $d\sigma_{1\,\,+,+,+,+}/d\sigma_{2\,\,+,+,+,+}$,
where $d\sigma_{1(2)\,\,+,+,+,+}$ -- $4$-boson vertex is
included } (isn't included) \label{p5}
\end{center}
\end{minipage}
\end{figure}
\end{center}

The  considered process is very important to study quartic-boson vertex (see
fig.\ref{p5}). The contribution of four-boson vertex to total cross
section is investigated specially. Obviously, processes of
$\gamma\gamma$-scattering are sensitive to the  four-boson vertex construction.
The experiments with polarized particles give wide set of 
additional  data for investigation of electroweak gauge
models.

\section{ The lowest-order correction to the $\gamma\gamma\rightarrow W^+W^-$ process}

$ $

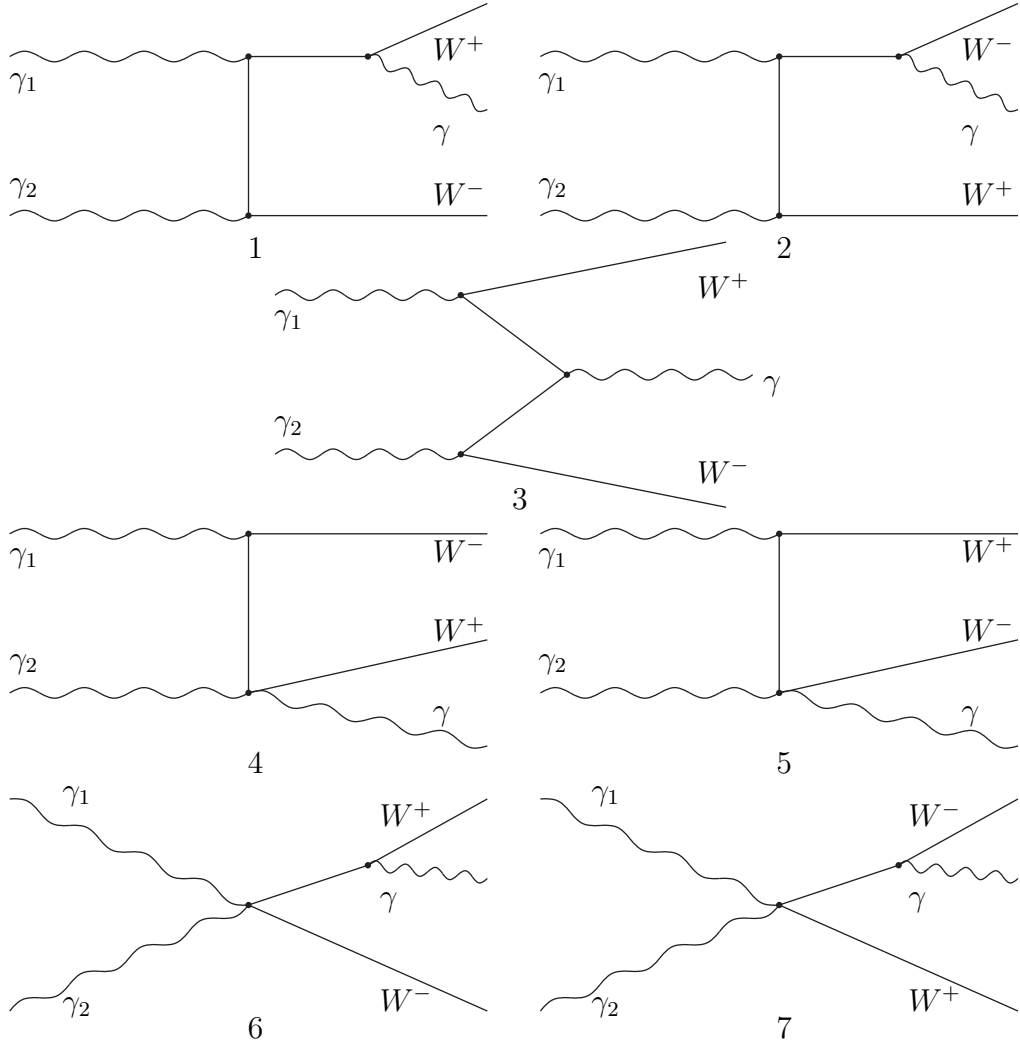
\begin{figure}[h!]
\begin{center}
\begin{picture}(400,400)(0,0)

\Photon(10,370)(100,370){2}{4}\put(10,360){$\gamma_1$}
\Photon(10,310)(100,310){2}{4}\put(10,320){$\gamma_2$}
\Vertex(100,370){1.2}
\Line(100,370)(145,370)
\Vertex(145,370){1.2}
\Line(145,370)(190,390)\put(170,370){$W^+$}
\Photon(145,370)(190,350){2}{4}\put(170,340){$\gamma$}
\Line(100,310)(100,370)
\Vertex(100,310){1.2}
\Line(100,310)(190,310)\put(170,315){$W^-$}
\put(100,295){$1$}

\Photon(210,370)(300,370){2}{4}\put(210,360){$\gamma_1$}
\Photon(210,310)(300,310){2}{4}\put(210,320){$\gamma_2$}
\Vertex(300,370){1.2}
\Line(300,370)(345,370)
\Vertex(345,370){1.2}
\Line(345,370)(390,390)\put(370,370){$W^-$}
\Photon(345,370)(390,350){2}{4}\put(370,340){$\gamma$}
\Line(300,310)(300,370)
\Vertex(300,310){1.2}
\Line(300,310)(390,310)\put(370,315){$W^+$}
\put(300,295){$2$}

\Photon(110,280)(180,280){2}{4}\put(110,270){$\gamma_1$}
\Photon(110,220)(180,220){2}{4}\put(110,230){$\gamma_2$}
\Vertex(180,280){1.2}
\Line(180,280)(280,300)\put(270,280){$W^+$}
\Line(180,280)(220,250)
\Vertex(220,250){1.2}
\Photon(220,250)(290,250){2}{4}\put(295,245){$\gamma$}
\Line(180,220)(220,250)
\Vertex(180,220){1.2}
\Line(180,220)(280,200)\put(270,210){$W^-$}
\put(200,200){$3$}

\Photon(10,190)(100,190){2}{4}\put(10,180){$\gamma_1$}
\Photon(10,130)(100,130){2}{4}\put(10,140){$\gamma_2$}
\Vertex(100,190){1.2}
\Line(100,190)(190,190)\put(170,180){$W^-$}
\Line(100,130)(100,190)
\Vertex(100,130){1.2}
\Line(100,130)(190,150)\put(170,150){$W^+$}
\Photon(100,130)(190,110){2}{4}\put(170,120){$\gamma$}
\put(100,100){$4$}

\Photon(210,190)(300,190){2}{4}\put(210,180){$\gamma_1$}
\Photon(210,130)(300,130){2}{4}\put(210,140){$\gamma_2$}
\Vertex(300,190){1.2}
\Line(300,190)(390,190)\put(370,180){$W^+$}
\Line(300,130)(300,190)
\Vertex(300,130){1.2}
\Line(300,130)(390,150)\put(370,150){$W^-$}
\Photon(300,130)(390,110){2}{4}\put(370,120){$\gamma$}
\put(300,100){$5$}

\Photon(10,90)(100,50){2}{4}\put(30,90){$\gamma_1$}
\Photon(10,10)(100,50){2}{4}\put(30,10){$\gamma_2$}
\Vertex(100,50){1.2}
\Line(100,50)(190,10)\put(150,10){$W^-$}
\Line(100,50)(145,65)
\Vertex(145,65){1.2}
\Line(145,65)(190,90)\put(150,80){$W^+$}
\Photon(145,65)(190,60){2}{4}\put(150,50){$\gamma$}
\put(100,0){$6$}

\Photon(210,90)(300,50){2}{4}\put(230,90){$\gamma_1$}
\Photon(210,10)(300,50){2}{4}\put(230,10){$\gamma_2$}
\Vertex(300,50){1.2}
\Line(300,50)(390,10)\put(350,10){$W^+$}
\Line(300,50)(345,65)
\Vertex(345,65){1.2}
\Line(345,65)(390,90)\put(350,80){$W^-$}
\Photon(345,65)(390,60){2}{4}\put(350,50){$\gamma$}
\put(300,0){$7$}

\end{picture}
\end{center}
\caption{The radiative corrections  for $\gamma\gamma\rightarrow
W^+W^-$ due to additional real photon}\label{p5x}
\end{figure}

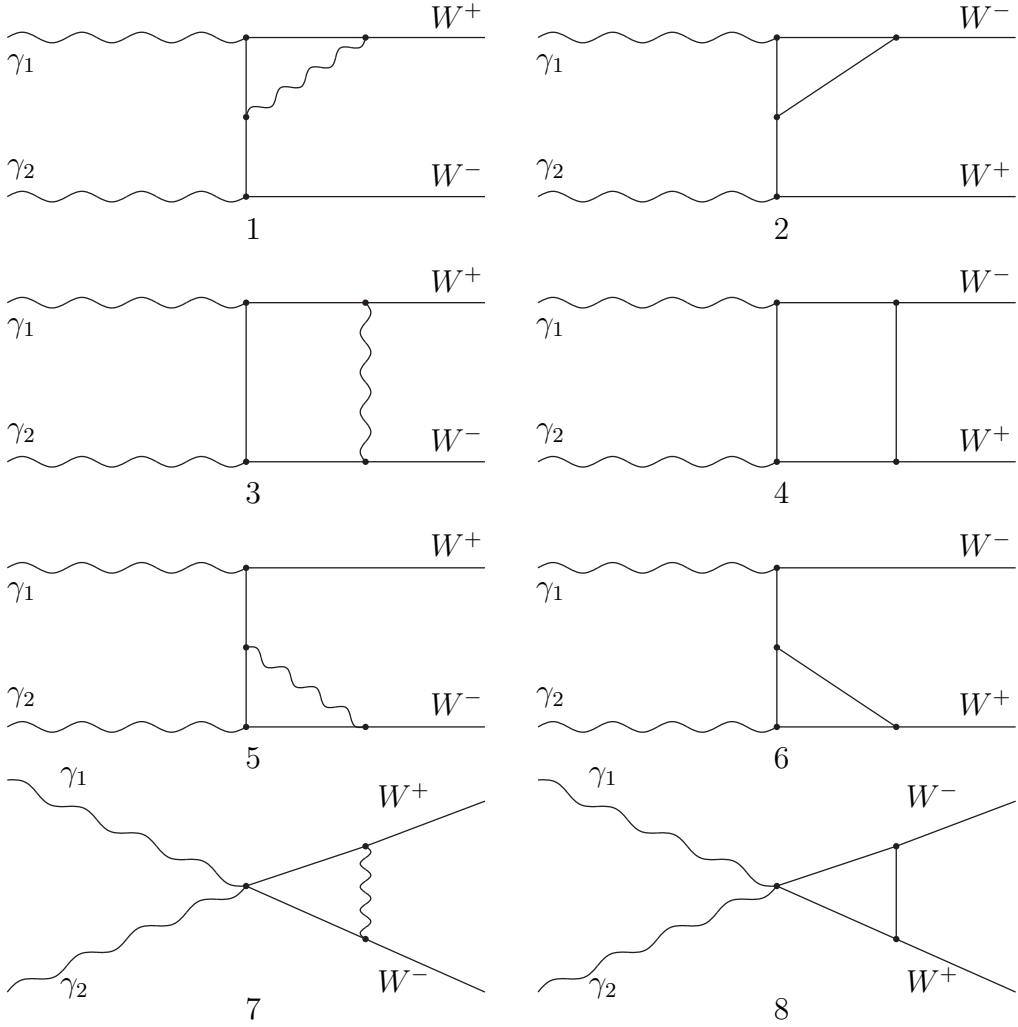
\begin{figure}[h!]
\begin{center}
\begin{picture}(400,400)(0,0)

\Photon(10,370)(100,370){2}{4}\put(10,360){$\gamma_1$}
\Photon(10,310)(100,310){2}{4}\put(10,320){$\gamma_2$}
\Vertex(100,370){1.2}
\Line(100,370)(190,370)\put(170,375){$W^+$}
\Vertex(145,370){1.2}
\Photon(145,370)(100,340){2}{4}
\Vertex(100,340){1.2}
\Line(100,310)(100,370)
\Vertex(100,310){1.2}
\Line(100,310)(190,310)\put(170,315){$W^-$}
\put(100,295){$1$}

\Photon(210,370)(300,370){2}{4}\put(210,360){$\gamma_1$}
\Photon(210,310)(300,310){2}{4}\put(210,320){$\gamma_2$}
\Vertex(300,370){1.2}
\Line(300,370)(390,370)\put(370,375){$W^-$}
\Vertex(345,370){1.2}
\Line(345,370)(300,340)
\Vertex(300,340){1.2}
\Line(300,310)(300,370)
\Vertex(300,310){1.2}
\Line(300,310)(390,310)\put(370,315){$W^+$}
\put(300,295){$2$}
\Photon(10,270)(100,270){2}{4}\put(10,260){$\gamma_1$}
\Photon(10,210)(100,210){2}{4}\put(10,220){$\gamma_2$}
\Vertex(100,270){1.2}
\Line(100,270)(190,270)\put(170,275){$W^+$}
\Vertex(145,270){1.2}
\Photon(145,270)(145,210){2}{4}
\Vertex(145,210){1.2}
\Line(100,210)(100,270)
\Vertex(100,210){1.2}
\Line(100,210)(190,210)\put(170,215){$W^-$}
\put(100,195){$3$}

\Photon(210,270)(300,270){2}{4}\put(210,260){$\gamma_1$}
\Photon(210,210)(300,210){2}{4}\put(210,220){$\gamma_2$}
\Vertex(300,270){1.2}
\Line(300,270)(390,270)\put(370,275){$W^-$}
\Vertex(345,270){1.2}
\Line(345,270)(345,210)
\Vertex(345,210){1.2}
\Line(300,210)(300,270)
\Vertex(300,210){1.2}
\Line(300,210)(390,210)\put(370,215){$W^+$}
\put(300,195){$4$}
\Photon(10,170)(100,170){2}{4}\put(10,160){$\gamma_1$}
\Photon(10,110)(100,110){2}{4}\put(10,120){$\gamma_2$}
\Vertex(100,170){1.2}
\Line(100,170)(190,170)\put(170,175){$W^+$}
\Vertex(145,110){1.2}
\Photon(145,110)(100,140){2}{4}
\Vertex(100,140){1.2}
\Line(100,110)(100,170)
\Vertex(100,110){1.2}
\Line(100,110)(190,110)\put(170,115){$W^-$}
\put(100,95){$5$}
\Photon(210,170)(300,170){2}{4}\put(210,160){$\gamma_1$}
\Photon(210,110)(300,110){2}{4}\put(210,120){$\gamma_2$}
\Vertex(300,170){1.2}
\Line(300,170)(390,170)\put(370,175){$W^-$}
\Vertex(345,110){1.2}
\Line(345,110)(300,140)
\Vertex(300,140){1.2}
\Line(300,110)(300,170)
\Vertex(300,110){1.2}
\Line(300,110)(390,110)\put(370,115){$W^+$}
\put(300,95){$6$}
\Photon(10,90)(100,50){2}{4}\put(30,90){$\gamma_1$}
\Photon(10,10)(100,50){2}{4}\put(30,10){$\gamma_2$}
\Vertex(100,50){1.2}
\Line(100,50)(190,10)\put(150,10){$W^-$}
\Line(100,50)(145,65)
\Vertex(145,65){1.2}
\Photon(145,65)(145,30){2}{4}
\Vertex(145,30){1.2}
\Line(145,65)(190,82)\put(150,80){$W^+$}
\put(100,0){$7$}
\Photon(210,90)(300,50){2}{4}\put(230,90){$\gamma_1$}
\Photon(210,10)(300,50){2}{4}\put(230,10){$\gamma_2$}
\Vertex(300,50){1.2}
\Line(300,50)(390,10)\put(350,10){$W^+$}
\Line(300,50)(345,65)
\Vertex(345,65){1.2}
\Line(345,65)(345,30)
\Vertex(345,30){1.2}
\Line(345,65)(390,82)\put(350,80){$W^-$}
\put(300,0){$8$}

\end{picture}
\end{center}
\caption{The radiative corrections  for $\gamma\gamma\rightarrow
W^+W^-$ due to additional virtual photon and Z-boson}\label{p6x}
\end{figure}

Obviously  the correct comparison of results for
these two processes  and precision analysis of data of the future
experiments are impossible without calculation of full set the
first-order radiative corrections, presented in figs. \ref{p5x},
 \ref{p6x} (see, for example, refs. \cite{bib4} -- \cite{bib9}).

As a rule,  the one-loop diagrams involving Higgs boson have been calculated
in order to study the possible investigation of the Higgs boson
via $\gamma\gamma \to H^* \to W^+W^-$.
See, for example, refs. \cite{bib01} -- \cite{bib03}.
Only the channels of longitudinal $W$-boson production
are sensitive to the Higgs mechanism, but the channels with $W$ 
transfer are insensitive.
This insensitivity to the Higgs sector render
$\gamma\gamma \to W^+W^-$ even more suitable
for the investigation of the self couplings. 

So, at $\sqrt{s} \sim 1000\,\, GeV$ the radiative correction
is about a value of Born cross section.
In this work we focus on the most important
QED one-loop corrections to $\gamma\gamma\to W^+W^-$.

\begin{figure}[h!]
 \leavevmode
\begin{minipage}[b]{.475\linewidth}
\centering
\includegraphics[width=\linewidth, height=3.8in, angle=0]{+-++.eps}
\caption{The differential cross section in Born approximation $d\sigma{+,-,+,+}/dy-$.}\label{p3}
\end{minipage}
\begin{minipage}[b]{.475\linewidth}
\centering
\includegraphics[width=\linewidth, height=3.8in, angle=0]{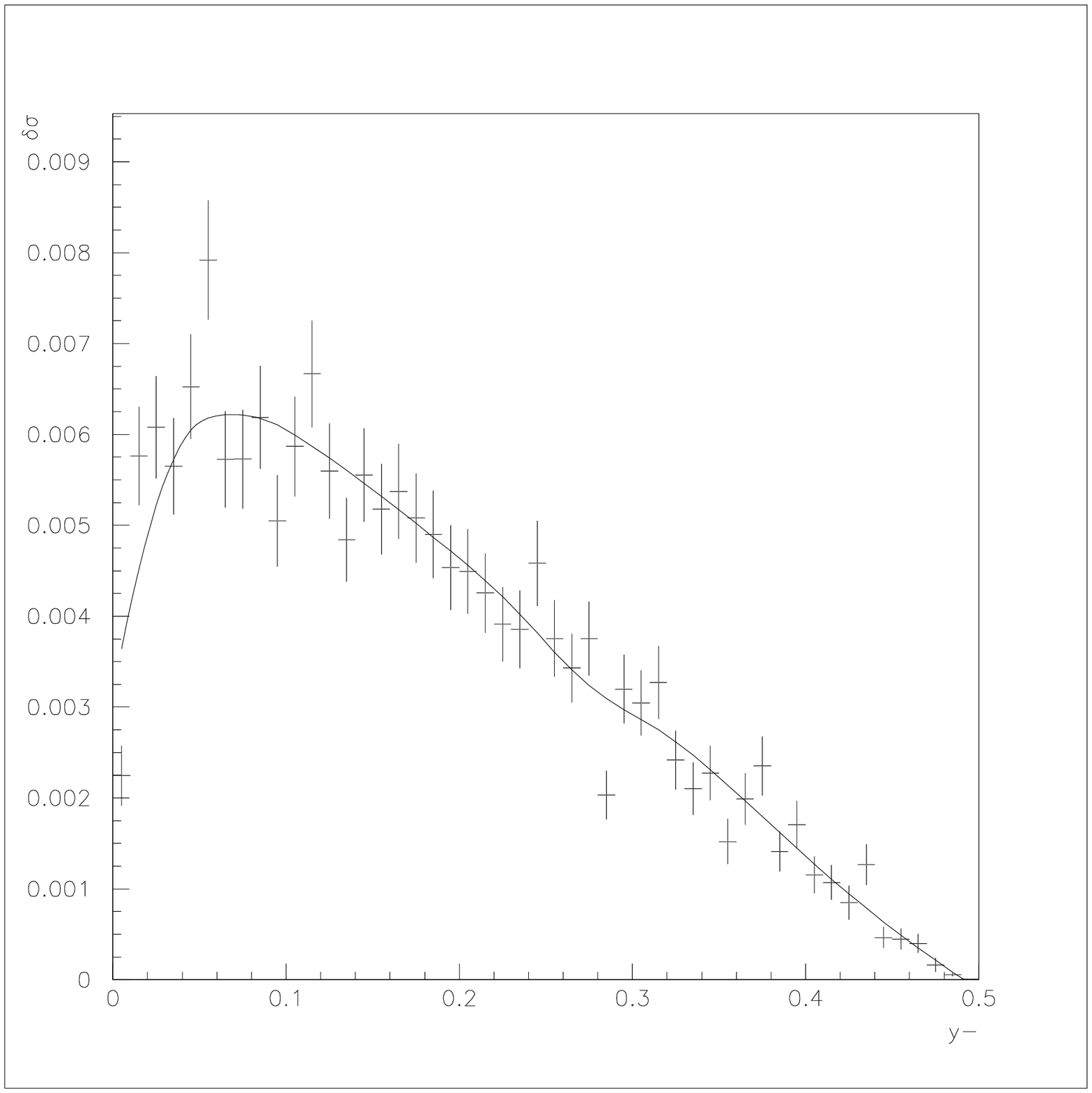}
\caption{The radiative correction for
$d\sigma{+,-,+,+}/dy-$.}\label{p2x}
\end{minipage}
\end{figure}

The additional real photon contribution
(see Feynman diagrams in fig. \ref{p5x})
and additional virtual photon contribution
(see Feynman diagrams in fig. \ref{p6x})
are shown in figs. \ref{p2x} -- \ref{p17}.


The radiative correction achives maximal value
in the kinematical region, where $y_-$ is near $0$.

There is a list of important expressions needed to calculate a
radiative contribution to the processes in figs. \ref{p5x} and \ref{p6x}.
Here
\begin{eqnarray}\label{rc3}
\begin{array}{l}
\displaystyle |M_1+M_2+M_3+M_4+M_5+M_6+M_7|^{2}_{soft} =
\left(\frac{A_1^{\gamma}+A_2^{\gamma}}{2(p_+k)} +
\frac{B_1^{\gamma}+B_2^{\gamma}}{2(p_-k)}\right)^2,
\end{array}
\end{eqnarray}
\begin{eqnarray}\label{rc1}
\begin{array}{l}
A^{\gamma}_1 
=C_{0\alpha\beta\mu\nu}[\Gamma_3^{\nu\beta\lambda}(k_2,-p_-,p_--k_2)
D_{\lambda\sigma}(p_--k_2)\Gamma_3^{\mu\sigma\rho}(k_1,k_2-p_,-p_+)(\delta_{\rho\eta}-
\\
\rule{0cm}{0.6cm} -p_{+\rho}p_{+\eta}/m_W^2)\Gamma_3^{\gamma\eta\alpha}(0,p_+,-p_+)+
\Gamma_3^{\gamma\rho\alpha}(0,-p_+,-P_+)(\delta_{\rho\eta} -
p_{+\rho}p_{+\eta}/m_W^2)\Gamma_4^{
\mu\nu\beta\eta}]_{\vec{p}_+=0},
\end{array}
\end{eqnarray}
\begin{eqnarray}\label{rc01}
\begin{array}{l}
A^{\gamma}_2  =
C_{0\alpha\beta\mu\nu}[\Gamma_3^{\mu\beta\lambda}(k_1,-p_-,p_--k_1)
D_{\lambda\sigma}(p_--k_1)\Gamma_3^{\nu\sigma\rho}(k_2,k_1-p_,-p_+)(\delta_{\rho\eta}-
\\
\rule{0cm}{0.6cm} -p_{+\rho}p_{+\eta}/m_W^2)\Gamma_3^{\gamma\eta\alpha}(0,p_+,-p_+)+
\Gamma_3^{\gamma\rho\alpha}(0,-p_+,-P_+)(\delta_{\rho\eta} -
p_{+\rho}p_{+\eta}/m_W^2)\Gamma_4^{
\mu\nu\beta\eta}]_{\vec{p}_+=0},
\end{array}
\end{eqnarray}
\begin{eqnarray}\label{rc2}
\begin{array}{l}
B^{\gamma}_1 
=C_{0\alpha\beta\mu\nu}[\Gamma_3^{\gamma\beta\eta}(0,-p_-,p_-)(\delta_{\rho\eta}-
p_{-\rho}p_{-\eta}/m_W^2)\Gamma_3^{\mu\rho\sigma}(k_1,-p_,-p_++k_2)D_{\lambda\sigma}(p_+-k_2)\times
\\
\rule{0cm}{0.6cm} \times\Gamma_3^{\nu\lambda\alpha}(k_2,p_+-k_2,-p_+)+
\Gamma_3^{\gamma\beta\rho}(0,-p_-,P_-)(\delta_{\rho\sigma} -
p_{-\rho}p_{-\sigma}/m_W^2)\Gamma_4^{
\mu\nu\alpha\sigma}]_{\vec{p}_+=0},\\ \rule{0cm}{0.6cm}B^{\gamma}_2 
=C_{0\alpha\beta\mu\nu}[\Gamma_3^{\gamma\beta\eta}(0,-p_-,p_-)(\delta_{\rho\eta}-
p_{-\rho}p_{-\eta}/m_W^2)\Gamma_3^{\nu\rho\sigma}(k_2,-p_,-p_++k_1)D_{\lambda\sigma}(p_+-k_1)\times\!\!\!\!\!\!\!\!
\\
 \rule{0cm}{0.6cm} \times\Gamma_3^{\mu\lambda\alpha}(k_1,p_+-k_1,-p_+)+
\Gamma_3^{\gamma\beta\rho}(0,-p_-,P_-)(\delta_{\rho\sigma} -
p_{-\rho}p_{-\sigma}/m_W^2)\Gamma_4^{
\mu\nu\alpha\sigma}]_{\vec{p}_+=0}, 
\end{array}
\end{eqnarray}
After that one can define the expressions of $D_1,D_2,D_3$.
\begin{eqnarray}\label{rc4}
\begin{array}{l}
D_1 =
(|A_1^{\gamma}+A_2^{\gamma}|^2p_0^2+|B_1^{\gamma}+B_2^{\gamma}|^2m_W^2+
2Re(A_1^{\gamma}+A_2^{\gamma}
(B_1^{\gamma}+B_2^{\gamma})p_-^0m)(8m_W^2p_-^{02})^{-1},
 \\ \rule{0cm}{0.6cm} D_2  =2
(|A_1^{\gamma}+A_2^{\gamma}|^2p_0^2+2Re(A_1^{\gamma}+A_2^{\gamma}
(B_1^{\gamma}+B_2^{\gamma})p_-^0m)(8m_W^2p_-^{02})^{-1},\\ \rule{0cm}{0.6cm}D_3  =
(|A_1^{\gamma}+A_2^{\gamma}|^2p_0^2)(8m_W^2p_-^{02})^{-1}.
\end{array}
\end{eqnarray}
Note, really it is impossible to obtain information with high precision from experimental data without the high-oder effects consideration \cite{bib9}. 
The radiative correction is defined as
\begin{eqnarray}\label{rc5}
\displaystyle \int(\sum_i M_i)^2\frac{d^3k}{(2\pi)^32k^0} = \int(\sum_i
M_i)^2\frac{d^{n-1}k}{(2\pi)^{n-1}2k} =
\int\frac{2\pi^{1/2n-1}k^{n-2}\sin^{n-3}{\theta}
dkd\theta}{\Gamma(1/2n-1)(2\pi)^{n-1}2k}(\sum_i M_i)^2,\qquad\!\!\!\!\!\!\!\!\!\!\!\!\!\!\!\nonumber\\ \rule{0cm}{0.8cm}
\displaystyle R.C. =
\frac{2\pi^{1/2n-1}}{(2\pi)^{n-1}\Gamma(1/2n-1)}\int_0^{\omega}k^{n-5}dk\int_0^{\pi}\sin^{n-3}{\theta}\frac{[D_1-v\cos{\theta}D_2
+ v^2\cos^2{\theta}D_3]}{(1-v\cos{\theta})^2},\qquad\quad\!\!
\end{eqnarray}
where $v=|\vec{p}_-|/p^0_-$. The first two terms in Loran`s series
are
\begin{eqnarray}\label{rc6}
\displaystyle C_{-1} =
\frac{1}{16\pi^2}\left[-\frac{2D_1}{1-v^2}-\frac{2D_3}{1-v^2}-\frac{2D_2}{(1-v^2)^2}-\frac{1}{v}\log{(\frac{1+v}{1-v})}[
D_2+2D_3]+2D_3\right], \rule{0cm}{0.8cm}\quad\,\,\nonumber\\ \displaystyle C_0 =
\frac{1}{16\pi^2}[(\log{\omega}-\log{2\sqrt{2}}+\frac{1}{2}C)(16\pi^2C_{-1})+\frac{1}{2}(\frac{4\log{2}}{1-v^2}+\frac{2}{1-v^2}\log{\frac{1-v}{1+v}})\times\rule{0cm}{0.8cm}\quad\,\,\\
\displaystyle \times(D_1-D_2+D_3)-\frac{1}{v}(\log{\frac{1-v}{1+v}}(\log{(1-\frac{1}{v})}+\log{(1+\frac{1}{v})})
+\Phi[\frac{1+v}{1-v}]-\Phi[\frac{1-v}{1+v}]\times\!\!\!\!\nonumber\\ \rule{0cm}{0.6cm}
\displaystyle\times(D_2-2D_3-2D_3(\log{2}-1)))].\qquad\qquad\qquad\qquad\qquad\qquad\qquad\qquad\quad\,\,\,\nonumber
\end{eqnarray}
There are some integrals needed for the calculation of a virtual
$\gamma$-contributions. This part of the correction was calculated
using integration package of the  $\sl Mathematica 4.0$.
\begin{eqnarray}\label{vc1}
\begin{array}{l}
\displaystyle I_1 =
\int\frac{d^nk}{(2\pi)^n}\frac{1}{k^2((p'-k)^2-m_W^2)((p+k)^2-m_W^2)}
= \\ \displaystyle \rule{0cm}{0.8cm} =
\frac{-i}{16\pi^2}\frac{1}{2}((\frac{1}{n-4}+\frac{1}{2}C-\log{2\sqrt{\pi}}+\log{m})-\frac{1}{2}(1-\log{2}-
\log{(1+\frac{pp'}{m_W^2})}))\frac{1}{m_W^2+pp'}, \\ \displaystyle \rule{0cm}{0.8cm}
I_{\alpha} =
\int\frac{d^nk}{(2\pi)^n}\frac{k_{\alpha}}{k^2((p'-k)^2-m_W^2)((p+k)^2-m_W^2)}
= \\ \displaystyle  \rule{0cm}{0.8cm}
=\frac{1}{q^2}\frac{1}{2p(p+p')/m_W^2+1}\left[(\frac{2p(p+p')}{m_W^2}+2)\log(\frac{2p(p+p')}{m_W^2}+2)-(\frac{2p(p+p')}{m_W^2}+1)\right]q_{\alpha},
\\ \displaystyle\rule{0cm}{0.8cm} I_{\alpha\beta} =
\int\frac{d^nk}{(2\pi)^n}\frac{k_{\alpha\beta}}{k^2((p'-k)^2-m_W^2)((p+k)^2-m_W^2)}
= \\ \displaystyle\rule{0cm}{0.8cm} = A\delta_{\alpha\beta}+B(p_{\alpha}p_{\beta} +
p'_{\alpha}p'_{\beta})-C(p_{\alpha}p'_{\beta}+p_{\beta}p'_{\alpha}),
\end{array}
\end{eqnarray}
where $A,B,C$ are given by following equations
\begin{eqnarray}\label{vc2}
\begin{array}{l}
\displaystyle A =\frac{L^{(n)}+2(pp')(B+C)}{4},\\ \displaystyle\rule{0cm}{0.8cm} B
=\frac{J_1+2(-m_W^2-2(pp'))C}{2(pp')}, \\ \displaystyle \rule{0cm}{0.8cm} C =
\frac{2J_2-L^{(n)}-J_1[-m_W^2/4-3/2(pp')]/(2(pp'))}{7/4 (pp')+(-m_W^2-2(pp'))/(pp')(-m_W^2/4+3/2(pp'))},
\end{array}
\end{eqnarray}
\begin{eqnarray}\label{vc3}
\begin{array}{l}
\displaystyle L^{(n)}
=\frac{-2i}{16\pi^2}[(\frac{1}{n-4}+\log{m_W}-\log{2\sqrt{\pi}}+\frac{1}{2}C)+\frac{1}{\chi}((\chi+1)\log(\chi+1)-\chi)],
\\ \displaystyle\rule{0cm}{0.8cm}  J_1 = D_0^{(n)} - D_1^{(n)} +
\frac{p^2-m_W^2}{q^2}\frac{1}{\chi}((\chi+1)\log(\chi+1)-\chi),
\\ \displaystyle\rule{0cm}{0.8cm} J_2  = D_2^{(n)} -
\frac{p^2-m_W^2}{q^2}\frac{1}{\chi}((\chi+1)\log(\chi+1)-\chi), \\ \displaystyle\rule{0cm}{0.8cm} 
D_0^{(n)} =\frac{-i}{16\pi^2}[(\frac{1}{n-4}+\frac{1}{2}C-1) +
\log{m_W}],\\ \displaystyle\rule{0cm}{0.8cm}  D_1^{(n)}
=\frac{-i}{16\pi^2}[(\frac{1}{n-4}+\log{m_W}-\log{2\sqrt{\pi}}+\frac{1}{2}C)+\frac{1}{\chi^2}(\frac{(\chi+1)^2}{2}\log(\chi+1)-
\\ \displaystyle\rule{0cm}{0.8cm} 
 -\frac{(\chi+1)^2}{4}+\frac{1}{4}-(\chi+1)\log(\chi+1)+\chi)], \\ \displaystyle\rule{0cm}{0.8cm} 
D_2^{(n)} =
\frac{-2i}{16\pi^2}[(\frac{1}{n-4}+\log{m_W}-\log{2\sqrt{\pi}}+\frac{1}{2}C)
+ \frac{1}{\chi}((\chi+1)\log(\chi+1)-\chi)- \\ \displaystyle\rule{0cm}{0.8cm} 
 -\frac{1}{2}(\frac{1}{\chi^2}(\frac{(\chi+1)^2}{2}\log(\chi+1)-\frac{(\chi+1)^2}{4}+\frac{1}{4}-(\chi+1)\log(\chi+1)+\chi))],\\ \displaystyle\rule{0cm}{0.8cm} 
\chi =2p(p+p')/m_W^2+1,
\end{array}
\end{eqnarray}
where $C$ - Euler's constant.

\newpage

\section*{Appendix A}
$ $

Using expressions (\ref{a1}) -- (\ref{b10}) for the Feynman diagrams figs. $1-3$  we have
obtained the following matrix element for $\gamma\gamma\rightarrow W^+W^-$:
\begin{eqnarray}\label{ap1}
\begin{array}{c}
M = G_v \sum_{i=1}^{3} M_i,
\end{array}
\end{eqnarray}  
where $M_i$ ($i=1,3$) are matrix elements related to the Feynman diagrams according to $i$.
\begin{eqnarray}\label{ap2}
\begin{array}{c}
\displaystyle M_1 = \frac{1}{(k_1-p_+)^2 - m^2}({p_-}.{{\epsilon }_2}\multsp {{\epsilon }_-}.{{\epsilon }_1}\multsp {p_+}.{{\epsilon }_+}-{p_-}.{{\epsilon }_1}\multsp {{\epsilon }_-}.{{\epsilon
}_2}\multsp {p_+}.{{\epsilon }_+}-2\multsp {p_-}.{{\epsilon }_2}\multsp {{\epsilon }_-}.{{\epsilon }_+}\multsp {p_+}.{{\epsilon }_1}+  \\
\displaystyle 
\hspace{1.em} 2\multsp {p_-}.{{\epsilon }_+}\multsp {{\epsilon }_-}.{{\epsilon }_2}\multsp {p_+}.{{\epsilon }_1}+\frac{{p_-}.{{\epsilon }_2}\multsp
{{\epsilon }_-}.{p_+}\multsp {p_+}.{{\epsilon }_+}\multsp {p_+}.{{\epsilon }_1}}{{m^2}}-  \\
\displaystyle 
\hspace{1.em} \frac{{p_-}.{{\epsilon }_2}\multsp {{\epsilon }_-}.{k_1}\multsp {p_+}.{{\epsilon }_+}\multsp {p_+}.{{\epsilon }_1}}{{m^2}}-\frac{{p_-}.{p_+}\multsp
{{\epsilon }_-}.{{\epsilon }_2}\multsp {p_+}.{{\epsilon }_+}\multsp {p_+}.{{\epsilon }_1}}{{m^2}}+  \\
\displaystyle 
\hspace{1.em} \frac{{p_-}.{k_1}\multsp {{\epsilon }_-}.{{\epsilon }_2}\multsp {p_+}.{{\epsilon }_+}\multsp {p_+}.{{\epsilon }_1}}{{m^2}}-\frac{{{\epsilon
}_-}.{{\epsilon }_2}\multsp {p_+}.{{\epsilon }_+}\multsp {p_+}.{k_2}\multsp {p_+}.{{\epsilon }_1}}{{m^2}}-  \\
\displaystyle 
\hspace{1.em} {{\epsilon }_-}.{{\epsilon }_1}\multsp {p_+}.{{\epsilon }_+}\multsp {p_+}.{{\epsilon }_2}+2\multsp {{\epsilon }_-}.{{\epsilon }_+}\multsp
{p_+}.{{\epsilon }_1}\multsp {p_+}.{{\epsilon }_2}+\frac{{{\epsilon }_-}.{k_2}\multsp {p_+}.{{\epsilon }_+}\multsp {p_+}.{{\epsilon }_1}\multsp {p_+}.{{\epsilon
}_2}}{{m^2}}-  \\
\displaystyle 
\hspace{1.em} 2\multsp {p_-}.{{\epsilon }_2}\multsp {{\epsilon }_-}.{{\epsilon }_1}\multsp {{\epsilon }_+}.{k_1}+2\multsp {p_-}.{{\epsilon }_1}\multsp
{{\epsilon }_-}.{{\epsilon }_2}\multsp {{\epsilon }_+}.{k_1}+2\multsp {{\epsilon }_-}.{{\epsilon }_1}\multsp {p_+}.{{\epsilon }_2}\multsp {{\epsilon
}_+}.{k_1}+  \\
\displaystyle 
\hspace{1.em} 2\multsp {{\epsilon }_-}.{{\epsilon }_2}\multsp {p_+}.{{\epsilon }_1}\multsp {{\epsilon }_+}.{k_2}+{p_-}.{{\epsilon }_2}\multsp {{\epsilon
}_-}.{p_+}\multsp {{\epsilon }_+}.{{\epsilon }_1}+{p_-}.{{\epsilon }_2}\multsp {{\epsilon }_-}.{k_1}\multsp {{\epsilon }_+}.{{\epsilon }_1}-  \\
\displaystyle 
\hspace{1.em} {p_-}.{p_+}\multsp {{\epsilon }_-}.{{\epsilon }_2}\multsp {{\epsilon }_+}.{{\epsilon }_1}-{p_-}.{k_1}\multsp {{\epsilon }_-}.{{\epsilon
}_2}\multsp {{\epsilon }_+}.{{\epsilon }_1}-\frac{{p_-}.{{\epsilon }_2}\multsp {{\epsilon }_-}.{p_+}\multsp {p_+}.{p_+}\multsp {{\epsilon }_+}.{{\epsilon
}_1}}{{m^2}}+  \\
\displaystyle 
\hspace{1.em} \frac{{p_-}.{{\epsilon }_2}\multsp {{\epsilon }_-}.{k_1}\multsp {p_+}.{p_+}\multsp {{\epsilon }_+}.{{\epsilon }_1}}{{m^2}}+\frac{{p_-}.{p_+}\multsp
{{\epsilon }_-}.{{\epsilon }_2}\multsp {p_+}.{p_+}\multsp {{\epsilon }_+}.{{\epsilon }_1}}{{m^2}}-  \\
\displaystyle 
\hspace{1.em} \frac{{p_-}.{k_1}\multsp {{\epsilon }_-}.{{\epsilon }_2}\multsp {p_+}.{p_+}\multsp {{\epsilon }_+}.{{\epsilon }_1}}{{m^2}}-{{\epsilon
}_-}.{{\epsilon }_2}\multsp {p_+}.{k_2}\multsp {{\epsilon }_+}.{{\epsilon }_1}+  \\
\displaystyle 
\hspace{1.em} \frac{{{\epsilon }_-}.{{\epsilon }_2}\multsp {p_+}.{p_+}\multsp {p_+}.{k_2}\multsp {{\epsilon }_+}.{{\epsilon }_1}}{{m^2}}-2\multsp
{{\epsilon }_-}.{k_1}\multsp {p_+}.{{\epsilon }_2}\multsp {{\epsilon }_+}.{{\epsilon }_1}+  \\
\displaystyle 
\hspace{1.em} {{\epsilon }_-}.{k_2}\multsp {p_+}.{{\epsilon }_2}\multsp {{\epsilon }_+}.{{\epsilon }_1}-\frac{{{\epsilon }_-}.{k_2}\multsp {p_+}.{p_+}\multsp
{p_+}.{{\epsilon }_2}\multsp {{\epsilon }_+}.{{\epsilon }_1}}{{m^2}}-  \\
\displaystyle 
\hspace{1.em} 2\multsp {{\epsilon }_-}.{p_+}\multsp {p_+}.{{\epsilon }_1}\multsp {{\epsilon }_+}.{{\epsilon }_2}+2\multsp {{\epsilon }_-}.{k_1}\multsp
{p_+}.{{\epsilon }_1}\multsp {{\epsilon }_+}.{{\epsilon }_2}-2\multsp {{\epsilon }_-}.{k_2}\multsp {p_+}.{{\epsilon }_1}\multsp {{\epsilon }_+}.{{\epsilon
}_2}+  \\
\displaystyle 
\hspace{1.em} \frac{{p_-}.{{\epsilon }_2}\multsp {{\epsilon }_-}.{p_+}\multsp {{\epsilon }_+}.{{\epsilon }_1}\multsp {k_1}.{k_1}}{{m^2}}-\frac{{p_-}.{{\epsilon
}_2}\multsp {{\epsilon }_-}.{k_1}\multsp {{\epsilon }_+}.{{\epsilon }_1}\multsp {k_1}.{k_1}}{{m^2}}-  \\
\displaystyle 
\hspace{1.em} \frac{{p_-}.{p_+}\multsp {{\epsilon }_-}.{{\epsilon }_2}\multsp {{\epsilon }_+}.{{\epsilon }_1}\multsp {k_1}.{k_1}}{{m^2}}+\frac{{p_-}.{k_1}\multsp
{{\epsilon }_-}.{{\epsilon }_2}\multsp {{\epsilon }_+}.{{\epsilon }_1}\multsp {k_1}.{k_1}}{{m^2}}-  \\
\displaystyle 
\hspace{1.em} \frac{{{\epsilon }_-}.{{\epsilon }_2}\multsp {p_+}.{k_2}\multsp {{\epsilon }_+}.{{\epsilon }_1}\multsp {k_1}.{k_1}}{{m^2}}+\frac{{{\epsilon
}_-}.{k_2}\multsp {p_+}.{{\epsilon }_2}\multsp {{\epsilon }_+}.{{\epsilon }_1}\multsp {k_1}.{k_1}}{{m^2}}+  \\
\displaystyle 
\hspace{1.em} \frac{{{\epsilon }_-}.{{\epsilon }_2}\multsp {p_+}.{{\epsilon }_+}\multsp {p_+}.{{\epsilon }_1}\multsp {k_1}.{k_2}}{{m^2}}-{{\epsilon
}_-}.{{\epsilon }_2}\multsp {{\epsilon }_+}.{{\epsilon }_1}\multsp {k_1}.{k_2}-  \\
\displaystyle 
\hspace{1.em} \frac{{{\epsilon }_-}.{{\epsilon }_2}\multsp {p_+}.{p_+}\multsp {{\epsilon }_+}.{{\epsilon }_1}\multsp {k_1}.{k_2}}{{m^2}}+\frac{{{\epsilon
}_-}.{{\epsilon }_2}\multsp {{\epsilon }_+}.{{\epsilon }_1}\multsp {k_1}.{k_1}\multsp {k_1}.{k_2}}{{m^2}}+  \\
\displaystyle 
\hspace{1.em} {p_-}.{{\epsilon }_2}\multsp {{\epsilon }_-}.{{\epsilon }_+}\multsp {k_1}.{{\epsilon }_1}-{p_-}.{{\epsilon }_+}\multsp {{\epsilon }_-}.{{\epsilon
}_2}\multsp {k_1}.{{\epsilon }_1}-{{\epsilon }_-}.{{\epsilon }_+}\multsp {p_+}.{{\epsilon }_2}\multsp {k_1}.{{\epsilon }_1}-  \\
\displaystyle 
\hspace{1.em} \frac{{p_-}.{{\epsilon }_2}\multsp {{\epsilon }_-}.{p_+}\multsp {{\epsilon }_+}.{k_1}\multsp {k_1}.{{\epsilon }_1}}{{m^2}}+\frac{{p_-}.{{\epsilon
}_2}\multsp {{\epsilon }_-}.{k_1}\multsp {{\epsilon }_+}.{k_1}\multsp {k_1}.{{\epsilon }_1}}{{m^2}}+  \\
\displaystyle 
\hspace{1.em} \frac{{p_-}.{p_+}\multsp {{\epsilon }_-}.{{\epsilon }_2}\multsp {{\epsilon }_+}.{k_1}\multsp {k_1}.{{\epsilon }_1}}{{m^2}}-\frac{{p_-}.{k_1}\multsp
{{\epsilon }_-}.{{\epsilon }_2}\multsp {{\epsilon }_+}.{k_1}\multsp {k_1}.{{\epsilon }_1}}{{m^2}}+  \\
\displaystyle 
\hspace{1.em} \frac{{{\epsilon }_-}.{{\epsilon }_2}\multsp {p_+}.{k_2}\multsp {{\epsilon }_+}.{k_1}\multsp {k_1}.{{\epsilon }_1}}{{m^2}}-\frac{{{\epsilon
}_-}.{k_2}\multsp {p_+}.{{\epsilon }_2}\multsp {{\epsilon }_+}.{k_1}\multsp {k_1}.{{\epsilon }_1}}{{m^2}}-  \\
\displaystyle 
\hspace{1.em} {{\epsilon }_-}.{{\epsilon }_2}\multsp {{\epsilon }_+}.{k_2}\multsp {k_1}.{{\epsilon }_1}+{{\epsilon }_-}.{p_+}\multsp {{\epsilon }_+}.{{\epsilon
}_2}\multsp {k_1}.{{\epsilon }_1}-{{\epsilon }_-}.{k_1}\multsp {{\epsilon }_+}.{{\epsilon }_2}\multsp {k_1}.{{\epsilon }_1}+  \\
\displaystyle 
\hspace{1.em} {{\epsilon }_-}.{k_2}\multsp {{\epsilon }_+}.{{\epsilon }_2}\multsp {k_1}.{{\epsilon }_1}-\frac{{{\epsilon }_-}.{{\epsilon }_2}\multsp
{{\epsilon }_+}.{k_1}\multsp {k_1}.{k_2}\multsp {k_1}.{{\epsilon }_1}}{{m^2}}+{{\epsilon }_-}.{{\epsilon }_1}\multsp {p_+}.{{\epsilon }_+}\multsp
{k_1}.{{\epsilon }_2}-  \\
\end{array}
\end{eqnarray}
\begin{eqnarray}\nonumber
\begin{array}{c}
\displaystyle 
\hspace{1.em} 2\multsp {{\epsilon }_-}.{{\epsilon }_+}\multsp {p_+}.{{\epsilon }_1}\multsp {k_1}.{{\epsilon }_2}-\frac{{{\epsilon }_-}.{k_2}\multsp
{p_+}.{{\epsilon }_+}\multsp {p_+}.{{\epsilon }_1}\multsp {k_1}.{{\epsilon }_2}}{{m^2}}-2\multsp {{\epsilon }_-}.{{\epsilon }_1}\multsp {{\epsilon
}_+}.{k_1}\multsp {k_1}.{{\epsilon }_2}+  \\
\displaystyle 
\hspace{1.em} 2\multsp {{\epsilon }_-}.{p_+}\multsp {{\epsilon }_+}.{{\epsilon }_1}\multsp {k_1}.{{\epsilon }_2}+{{\epsilon }_-}.{k_2}\multsp {{\epsilon
}_+}.{{\epsilon }_1}\multsp {k_1}.{{\epsilon }_2}+\frac{{{\epsilon }_-}.{k_2}\multsp {p_+}.{p_+}\multsp {{\epsilon }_+}.{{\epsilon }_1}\multsp {k_1}.{{\epsilon
}_2}}{{m^2}}-  \\
\displaystyle 
\hspace{1.em} \frac{{{\epsilon }_-}.{k_2}\multsp {{\epsilon }_+}.{{\epsilon }_1}\multsp {k_1}.{k_1}\multsp {k_1}.{{\epsilon }_2}}{{m^2}}+{{\epsilon
}_-}.{{\epsilon }_+}\multsp {k_1}.{{\epsilon }_1}\multsp {k_1}.{{\epsilon }_2}+  \\
\displaystyle 
\hspace{1.em} \frac{{{\epsilon }_-}.{k_2}\multsp {{\epsilon }_+}.{k_1}\multsp {k_1}.{{\epsilon }_1}\multsp {k_1}.{{\epsilon }_2}}{{m^2}}-{{\epsilon
}_-}.{{\epsilon }_2}\multsp {p_+}.{{\epsilon }_+}\multsp {k_2}.{{\epsilon }_1}+2\multsp {{\epsilon }_-}.{{\epsilon }_2}\multsp {{\epsilon }_+}.{k_1}\multsp
{k_2}.{{\epsilon }_1}+  \\
\displaystyle 
\hspace{1.em} {{\epsilon }_-}.{p_+}\multsp {p_+}.{{\epsilon }_+}\multsp {{\epsilon }_1}.{{\epsilon }_2}-{{\epsilon }_-}.{k_1}\multsp {p_+}.{{\epsilon
}_+}\multsp {{\epsilon }_1}.{{\epsilon }_2}+{{\epsilon }_-}.{k_2}\multsp {p_+}.{{\epsilon }_+}\multsp {{\epsilon }_1}.{{\epsilon }_2}-  \\
\displaystyle 
\hspace{1.em} 2\multsp {{\epsilon }_-}.{p_+}\multsp {{\epsilon }_+}.{k_1}\multsp {{\epsilon }_1}.{{\epsilon }_2}+2\multsp {{\epsilon }_-}.{k_1}\multsp
{{\epsilon }_+}.{k_1}\multsp {{\epsilon }_1}.{{\epsilon }_2}-2\multsp {{\epsilon }_-}.{k_2}\multsp {{\epsilon }_+}.{k_1}\multsp {{\epsilon }_1}.{{\epsilon
}_2}),
\end{array}
\end{eqnarray}
\begin{eqnarray}\label{ap3}
\begin{array}{c}
\displaystyle M_2 = \frac{1}{(k_1-p_-)^2 - m^2}(2\multsp {p_-}.{{\epsilon }_1}\multsp {p_-}.{{\epsilon }_2}\multsp {{\epsilon }_-}.{{\epsilon }_+}-2\multsp {p_-}.{{\epsilon }_+}\multsp {p_-}.{{\epsilon
}_1}\multsp {{\epsilon }_-}.{{\epsilon }_2}+\frac{{p_-}.{{\epsilon }_-}\multsp {p_-}.{{\epsilon }_+}\multsp {p_-}.{{\epsilon }_1}\multsp {p_+}.{{\epsilon
}_2}}{{m^2}}-  \\
\displaystyle 
\hspace{1.em} 2\multsp {p_-}.{{\epsilon }_1}\multsp {{\epsilon }_-}.{{\epsilon }_+}\multsp {p_+}.{{\epsilon }_2}+{p_-}.{{\epsilon }_+}\multsp {{\epsilon
}_-}.{{\epsilon }_1}\multsp {p_+}.{{\epsilon }_2}-\frac{{p_-}.{p_-}\multsp {p_-}.{{\epsilon }_+}\multsp {{\epsilon }_-}.{{\epsilon }_1}\multsp {p_+}.{{\epsilon
}_2}}{{m^2}}-  \\
\displaystyle 
\hspace{1.em} 2\multsp {p_-}.{{\epsilon }_2}\multsp {{\epsilon }_-}.{{\epsilon }_1}\multsp {{\epsilon }_+}.{k_1}+2\multsp {p_-}.{{\epsilon }_1}\multsp
{{\epsilon }_-}.{{\epsilon }_2}\multsp {{\epsilon }_+}.{k_1}-\frac{{p_-}.{{\epsilon }_-}\multsp {p_-}.{{\epsilon }_1}\multsp {p_+}.{{\epsilon }_2}\multsp
{{\epsilon }_+}.{k_1}}{{m^2}}+  \\
\displaystyle 
\hspace{1.em} {{\epsilon }_-}.{{\epsilon }_1}\multsp {p_+}.{{\epsilon }_2}\multsp {{\epsilon }_+}.{k_1}+\frac{{p_-}.{p_-}\multsp {{\epsilon }_-}.{{\epsilon
}_1}\multsp {p_+}.{{\epsilon }_2}\multsp {{\epsilon }_+}.{k_1}}{{m^2}}+\frac{{p_-}.{{\epsilon }_-}\multsp {p_-}.{{\epsilon }_1}\multsp {p_-}.{{\epsilon
}_2}\multsp {{\epsilon }_+}.{k_2}}{{m^2}}+  \\
\displaystyle 
\hspace{1.em} {p_-}.{{\epsilon }_2}\multsp {{\epsilon }_-}.{{\epsilon }_1}\multsp {{\epsilon }_+}.{k_2}-\frac{{p_-}.{p_-}\multsp {p_-}.{{\epsilon
}_2}\multsp {{\epsilon }_-}.{{\epsilon }_1}\multsp {{\epsilon }_+}.{k_2}}{{m^2}}-2\multsp {p_-}.{{\epsilon }_1}\multsp {{\epsilon }_-}.{{\epsilon
}_2}\multsp {{\epsilon }_+}.{k_2}-  \\
\displaystyle 
\hspace{1.em} {p_-}.{{\epsilon }_-}\multsp {p_-}.{{\epsilon }_2}\multsp {{\epsilon }_+}.{{\epsilon }_1}+2\multsp {p_-}.{{\epsilon }_2}\multsp {{\epsilon
}_-}.{k_1}\multsp {{\epsilon }_+}.{{\epsilon }_1}+{p_-}.{{\epsilon }_-}\multsp {p_+}.{{\epsilon }_2}\multsp {{\epsilon }_+}.{{\epsilon }_1}-  \\
\displaystyle 
\hspace{1.em} 2\multsp {{\epsilon }_-}.{k_1}\multsp {p_+}.{{\epsilon }_2}\multsp {{\epsilon }_+}.{{\epsilon }_1}-\frac{{p_-}.{{\epsilon }_-}\multsp
{p_-}.{p_+}\multsp {p_-}.{{\epsilon }_1}\multsp {{\epsilon }_+}.{{\epsilon }_2}}{{m^2}}-\frac{{p_-}.{{\epsilon }_-}\multsp {p_-}.{k_2}\multsp {p_-}.{{\epsilon
}_1}\multsp {{\epsilon }_+}.{{\epsilon }_2}}{{m^2}}+  \\
\displaystyle 
\hspace{1.em} 2\multsp {p_-}.{{\epsilon }_1}\multsp {{\epsilon }_-}.{p_+}\multsp {{\epsilon }_+}.{{\epsilon }_2}+2\multsp {p_-}.{{\epsilon }_1}\multsp
{{\epsilon }_-}.{k_2}\multsp {{\epsilon }_+}.{{\epsilon }_2}-{p_-}.{p_+}\multsp {{\epsilon }_-}.{{\epsilon }_1}\multsp {{\epsilon }_+}.{{\epsilon
}_2}+  \\
\displaystyle 
\hspace{1.em} \frac{{p_-}.{p_-}\multsp {p_-}.{p_+}\multsp {{\epsilon }_-}.{{\epsilon }_1}\multsp {{\epsilon }_+}.{{\epsilon }_2}}{{m^2}}-{p_-}.{k_2}\multsp
{{\epsilon }_-}.{{\epsilon }_1}\multsp {{\epsilon }_+}.{{\epsilon }_2}+\frac{{p_-}.{p_-}\multsp {p_-}.{k_2}\multsp {{\epsilon }_-}.{{\epsilon }_1}\multsp
{{\epsilon }_+}.{{\epsilon }_2}}{{m^2}}+  \\
\displaystyle 
\hspace{1.em} \frac{{p_-}.{{\epsilon }_-}\multsp {p_-}.{{\epsilon }_1}\multsp {p_+}.{k_1}\multsp {{\epsilon }_+}.{{\epsilon }_2}}{{m^2}}-{{\epsilon
}_-}.{{\epsilon }_1}\multsp {p_+}.{k_1}\multsp {{\epsilon }_+}.{{\epsilon }_2}-\frac{{p_-}.{p_-}\multsp {{\epsilon }_-}.{{\epsilon }_1}\multsp {p_+}.{k_1}\multsp
{{\epsilon }_+}.{{\epsilon }_2}}{{m^2}}-  \\
\displaystyle 
\hspace{1.em} {p_-}.{{\epsilon }_-}\multsp {p_+}.{{\epsilon }_1}\multsp {{\epsilon }_+}.{{\epsilon }_2}+2\multsp {{\epsilon }_-}.{k_1}\multsp {p_+}.{{\epsilon
}_1}\multsp {{\epsilon }_+}.{{\epsilon }_2}+\frac{{p_-}.{{\epsilon }_+}\multsp {{\epsilon }_-}.{{\epsilon }_1}\multsp {p_+}.{{\epsilon }_2}\multsp
{k_1}.{k_1}}{{m^2}}-  \\
\displaystyle 
\hspace{1.em} \frac{{{\epsilon }_-}.{{\epsilon }_1}\multsp {p_+}.{{\epsilon }_2}\multsp {{\epsilon }_+}.{k_1}\multsp {k_1}.{k_1}}{{m^2}}+\frac{{p_-}.{{\epsilon
}_2}\multsp {{\epsilon }_-}.{{\epsilon }_1}\multsp {{\epsilon }_+}.{k_2}\multsp {k_1}.{k_1}}{{m^2}}-  \\
\displaystyle 
\hspace{1.em} \frac{{p_-}.{p_+}\multsp {{\epsilon }_-}.{{\epsilon }_1}\multsp {{\epsilon }_+}.{{\epsilon }_2}\multsp {k_1}.{k_1}}{{m^2}}-\frac{{p_-}.{k_2}\multsp
{{\epsilon }_-}.{{\epsilon }_1}\multsp {{\epsilon }_+}.{{\epsilon }_2}\multsp {k_1}.{k_1}}{{m^2}}+\frac{{{\epsilon }_-}.{{\epsilon }_1}\multsp {p_+}.{k_1}\multsp
{{\epsilon }_+}.{{\epsilon }_2}\multsp {k_1}.{k_1}}{{m^2}}+  \\
\displaystyle 
\hspace{1.em} \frac{{p_-}.{{\epsilon }_-}\multsp {p_-}.{{\epsilon }_1}\multsp {{\epsilon }_+}.{{\epsilon }_2}\multsp {k_1}.{k_2}}{{m^2}}-{{\epsilon
}_-}.{{\epsilon }_1}\multsp {{\epsilon }_+}.{{\epsilon }_2}\multsp {k_1}.{k_2}-\frac{{p_-}.{p_-}\multsp {{\epsilon }_-}.{{\epsilon }_1}\multsp {{\epsilon
}_+}.{{\epsilon }_2}\multsp {k_1}.{k_2}}{{m^2}}+  \\
\displaystyle 
\hspace{1.em} \frac{{{\epsilon }_-}.{{\epsilon }_1}\multsp {{\epsilon }_+}.{{\epsilon }_2}\multsp {k_1}.{k_1}\multsp {k_1}.{k_2}}{{m^2}}-{p_-}.{{\epsilon
}_2}\multsp {{\epsilon }_-}.{{\epsilon }_+}\multsp {k_1}.{{\epsilon }_1}+{p_-}.{{\epsilon }_+}\multsp {{\epsilon }_-}.{{\epsilon }_2}\multsp {k_1}.{{\epsilon
}_1}+  \\
\displaystyle 
\hspace{1.em} {{\epsilon }_-}.{{\epsilon }_+}\multsp {p_+}.{{\epsilon }_2}\multsp {k_1}.{{\epsilon }_1}-\frac{{p_-}.{{\epsilon }_+}\multsp {{\epsilon
}_-}.{k_1}\multsp {p_+}.{{\epsilon }_2}\multsp {k_1}.{{\epsilon }_1}}{{m^2}}-{{\epsilon }_-}.{{\epsilon }_2}\multsp {{\epsilon }_+}.{k_1}\multsp
{k_1}.{{\epsilon }_1}+  \\
\displaystyle 
\hspace{1.em} \frac{{{\epsilon }_-}.{k_1}\multsp {p_+}.{{\epsilon }_2}\multsp {{\epsilon }_+}.{k_1}\multsp {k_1}.{{\epsilon }_1}}{{m^2}}-\frac{{p_-}.{{\epsilon
}_2}\multsp {{\epsilon }_-}.{k_1}\multsp {{\epsilon }_+}.{k_2}\multsp {k_1}.{{\epsilon }_1}}{{m^2}}+{{\epsilon }_-}.{{\epsilon }_2}\multsp {{\epsilon
}_+}.{k_2}\multsp {k_1}.{{\epsilon }_1}-  \\
\displaystyle 
\hspace{1.em} {{\epsilon }_-}.{p_+}\multsp {{\epsilon }_+}.{{\epsilon }_2}\multsp {k_1}.{{\epsilon }_1}+\frac{{p_-}.{p_+}\multsp {{\epsilon }_-}.{k_1}\multsp
{{\epsilon }_+}.{{\epsilon }_2}\multsp {k_1}.{{\epsilon }_1}}{{m^2}}+\frac{{p_-}.{k_2}\multsp {{\epsilon }_-}.{k_1}\multsp {{\epsilon }_+}.{{\epsilon
}_2}\multsp {k_1}.{{\epsilon }_1}}{{m^2}}-  \\
\displaystyle 
\hspace{1.em} {{\epsilon }_-}.{k_2}\multsp {{\epsilon }_+}.{{\epsilon }_2}\multsp {k_1}.{{\epsilon }_1}-\frac{{{\epsilon }_-}.{k_1}\multsp {p_+}.{k_1}\multsp
{{\epsilon }_+}.{{\epsilon }_2}\multsp {k_1}.{{\epsilon }_1}}{{m^2}}-\frac{{{\epsilon }_-}.{k_1}\multsp {{\epsilon }_+}.{{\epsilon }_2}\multsp {k_1}.{k_2}\multsp
{k_1}.{{\epsilon }_1}}{{m^2}}-  \\
\displaystyle 
\hspace{1.em} 2\multsp {p_-}.{{\epsilon }_1}\multsp {{\epsilon }_-}.{{\epsilon }_+}\multsp {k_1}.{{\epsilon }_2}+2\multsp {p_-}.{{\epsilon }_+}\multsp
{{\epsilon }_-}.{{\epsilon }_1}\multsp {k_1}.{{\epsilon }_2}-\frac{{p_-}.{{\epsilon }_-}\multsp {p_-}.{{\epsilon }_1}\multsp {{\epsilon }_+}.{k_2}\multsp
{k_1}.{{\epsilon }_2}}{{m^2}}+  \\
\displaystyle 
\hspace{1.em} {{\epsilon }_-}.{{\epsilon }_1}\multsp {{\epsilon }_+}.{k_2}\multsp {k_1}.{{\epsilon }_2}+\frac{{p_-}.{p_-}\multsp {{\epsilon }_-}.{{\epsilon
}_1}\multsp {{\epsilon }_+}.{k_2}\multsp {k_1}.{{\epsilon }_2}}{{m^2}}+{p_-}.{{\epsilon }_-}\multsp {{\epsilon }_+}.{{\epsilon }_1}\multsp {k_1}.{{\epsilon
}_2}-  \\
\displaystyle 
\hspace{1.em} 2\multsp {{\epsilon }_-}.{k_1}\multsp {{\epsilon }_+}.{{\epsilon }_1}\multsp {k_1}.{{\epsilon }_2}-\frac{{{\epsilon }_-}.{{\epsilon
}_1}\multsp {{\epsilon }_+}.{k_2}\multsp {k_1}.{k_1}\multsp {k_1}.{{\epsilon }_2}}{{m^2}}+{{\epsilon }_-}.{{\epsilon }_+}\multsp {k_1}.{{\epsilon
}_1}\multsp {k_1}.{{\epsilon }_2}+  \\
\end{array}
\end{eqnarray}
\begin{eqnarray}\nonumber
\begin{array}{c}
\displaystyle 
\hspace{1.em} \frac{{{\epsilon }_-}.{k_1}\multsp {{\epsilon }_+}.{k_2}\multsp {k_1}.{{\epsilon }_1}\multsp {k_1}.{{\epsilon }_2}}{{m^2}}-{p_-}.{{\epsilon
}_-}\multsp {{\epsilon }_+}.{{\epsilon }_2}\multsp {k_2}.{{\epsilon }_1}+2\multsp {{\epsilon }_-}.{k_1}\multsp {{\epsilon }_+}.{{\epsilon }_2}\multsp
{k_2}.{{\epsilon }_1}+  \\
\displaystyle 
\hspace{1.em} {p_-}.{{\epsilon }_-}\multsp {p_-}.{{\epsilon }_+}\multsp {{\epsilon }_1}.{{\epsilon }_2}-2\multsp {p_-}.{{\epsilon }_+}\multsp {{\epsilon
}_-}.{k_1}\multsp {{\epsilon }_1}.{{\epsilon }_2}-{p_-}.{{\epsilon }_-}\multsp {{\epsilon }_+}.{k_1}\multsp {{\epsilon }_1}.{{\epsilon }_2}+  \\
\displaystyle 
\hspace{1.em} 2\multsp {{\epsilon }_-}.{k_1}\multsp {{\epsilon }_+}.{k_1}\multsp {{\epsilon }_1}.{{\epsilon }_2}+{p_-}.{{\epsilon }_-}\multsp {{\epsilon
}_+}.{k_2}\multsp {{\epsilon }_1}.{{\epsilon }_2}-2\multsp {{\epsilon }_-}.{k_1}\multsp {{\epsilon }_+}.{k_2}\multsp {{\epsilon }_1}.{{\epsilon }_2}),
\end{array}
\end{eqnarray}
\begin{eqnarray}\label{ap4}
\begin{array}{c}
\displaystyle M_3 = -{{\epsilon }_-}.{{\epsilon }_2}\multsp {{\epsilon }_+}.{{\epsilon }_1}-{{\epsilon }_-}.{{\epsilon }_1}\multsp {{\epsilon }_+}.{{\epsilon
}_2}+2\multsp {{\epsilon }_-}.{{\epsilon }_+}\multsp {{\epsilon }_1}.{{\epsilon }_2}
\end{array}
\end{eqnarray}
$k_1$, $k_2$, $p_{+}$, $p_{-}$ denote four-momenta of  $\gamma$, $\gamma$, $W^{+}$, $W^{-}$, respectively, 
and $\epsilon_1, \epsilon_2,\epsilon_{+}, \epsilon_{-}$ -- cooresponding polarizations.
\newpage
\section*{Appendix B}
\vspace{-10pt}
\newpage
\begin{center}
\begin{figure}[h!]
 \leavevmode
\begin{minipage}[b]{1.\linewidth}
\centering
\includegraphics[width=0.5\linewidth, height=3.8in, angle=0]{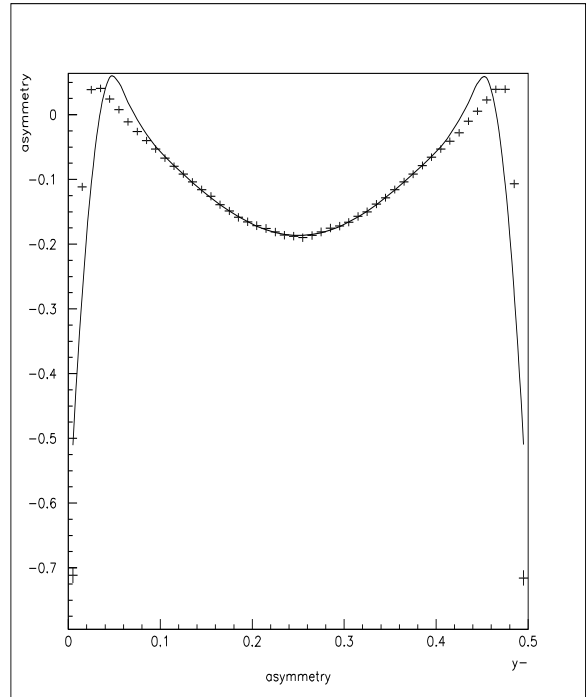}
\caption{The asymmetry $A_1$}\label{p9}
\end{minipage}
\end{figure}
\end{center}
\begin{figure}[h!]
 \leavevmode
\begin{minipage}[b]{.475\linewidth}
\centering
\includegraphics[width=\linewidth, height=3.8in, angle=0]{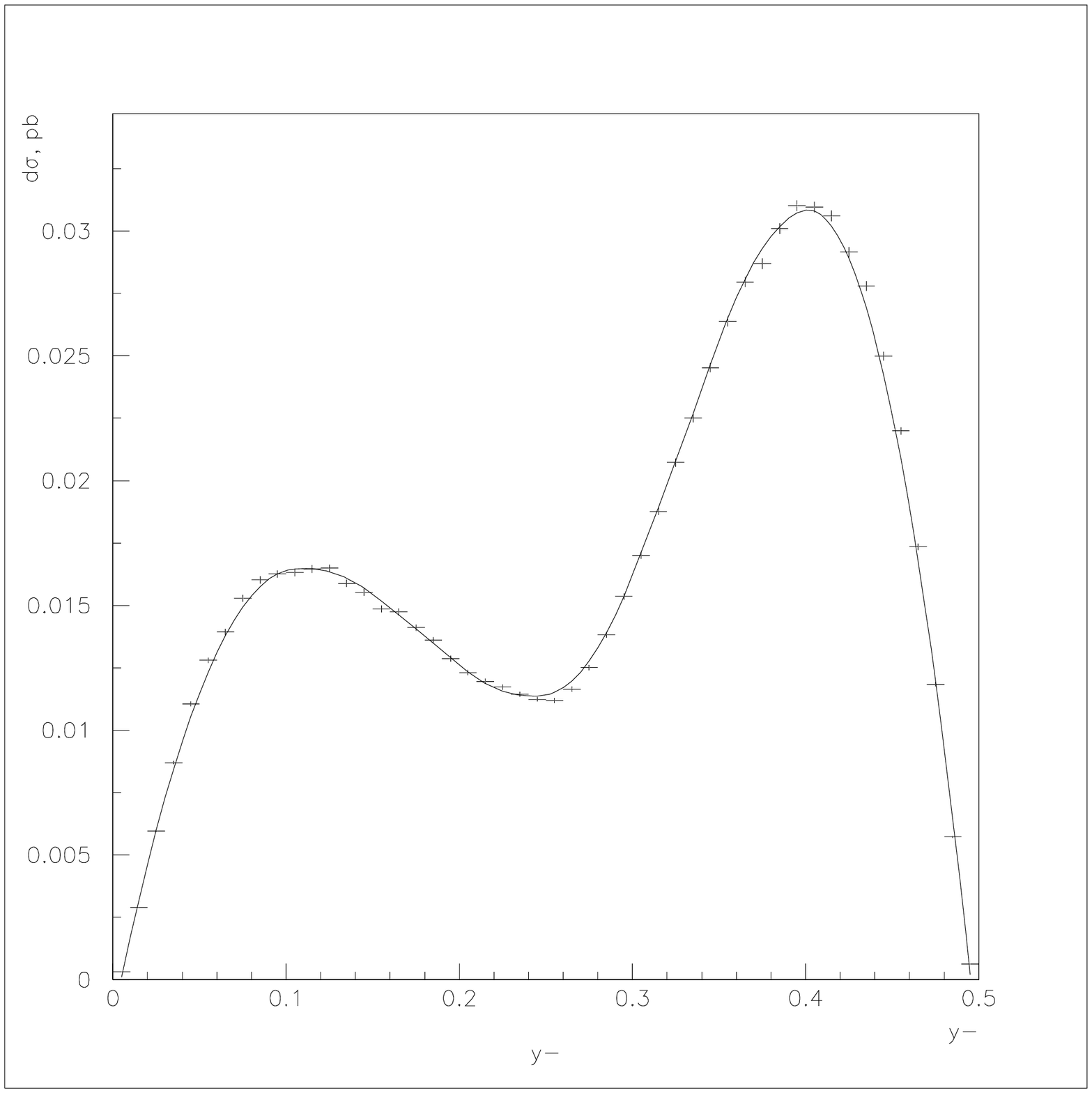}
\caption{The differential cross section $d\sigma_{+,+,+,-}/dy_-$}\label{p13}
\end{minipage}\hfill
\begin{minipage}[b]{.475\linewidth}
\centering
\includegraphics[width=\linewidth, height=3.8in, angle=0]{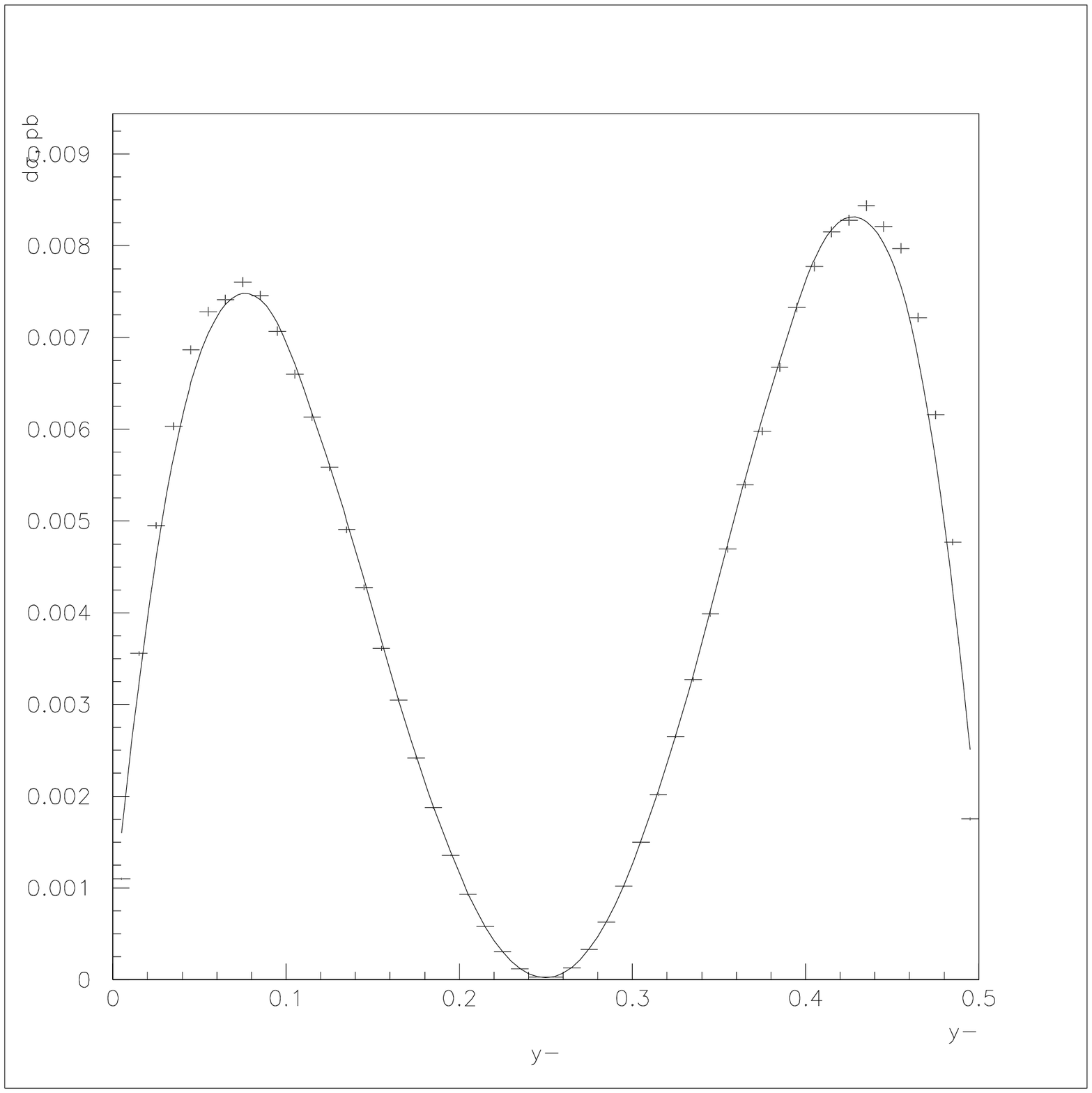}
\caption{The differential cross section $d\sigma_{+,-,+,-}/dy_-$}\label{p14}
\end{minipage}
\end{figure}

\newpage 
\begin{center}
\begin{figure}[h!]
\begin{minipage}[b]{1.\linewidth}
\centering
\includegraphics[width=0.5\linewidth, height=3.8in, angle=0]{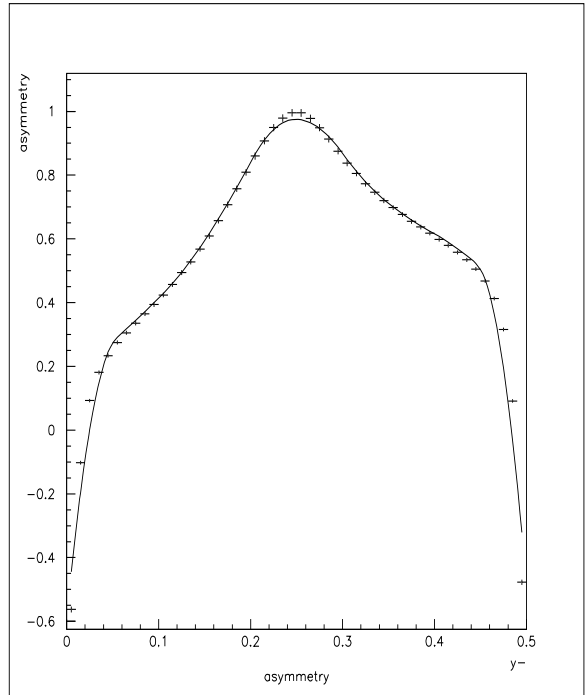}
\caption{The asymmetry $A_2$}\label{p15}
\end{minipage}
\end{figure}
\end{center}
\begin{figure}[h!]
 \leavevmode
\begin{minipage}[b]{.475\linewidth}
\centering
\includegraphics[width=\linewidth, height=3.8in, angle=0]{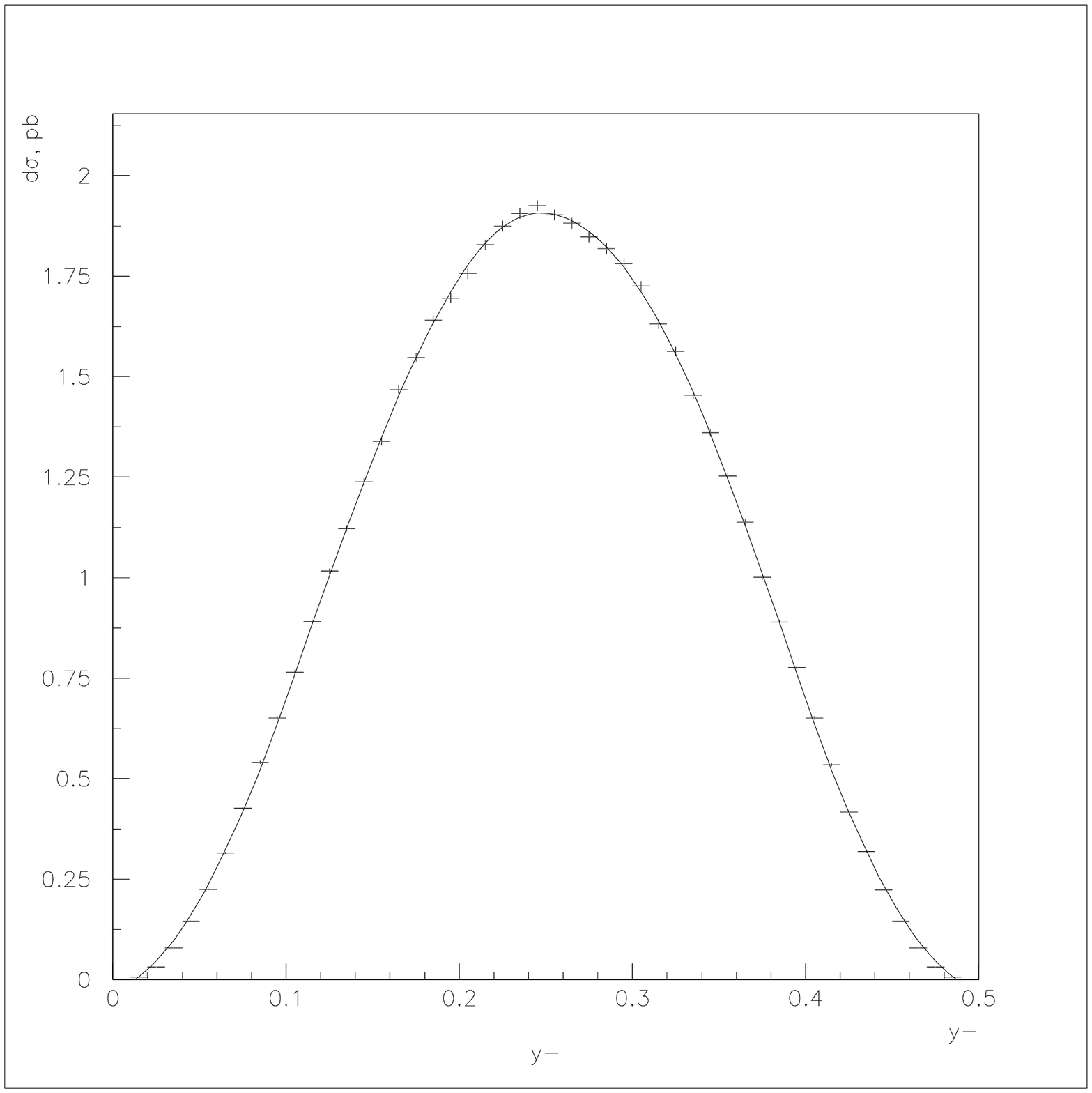}
\caption{The differential cross section $d\sigma_{+,+,0,0}/dy_-$}\label{p16}
\end{minipage}\hfill
\begin{minipage}[b]{.475\linewidth}
\centering
\includegraphics[width=\linewidth, height=3.8in, angle=0]{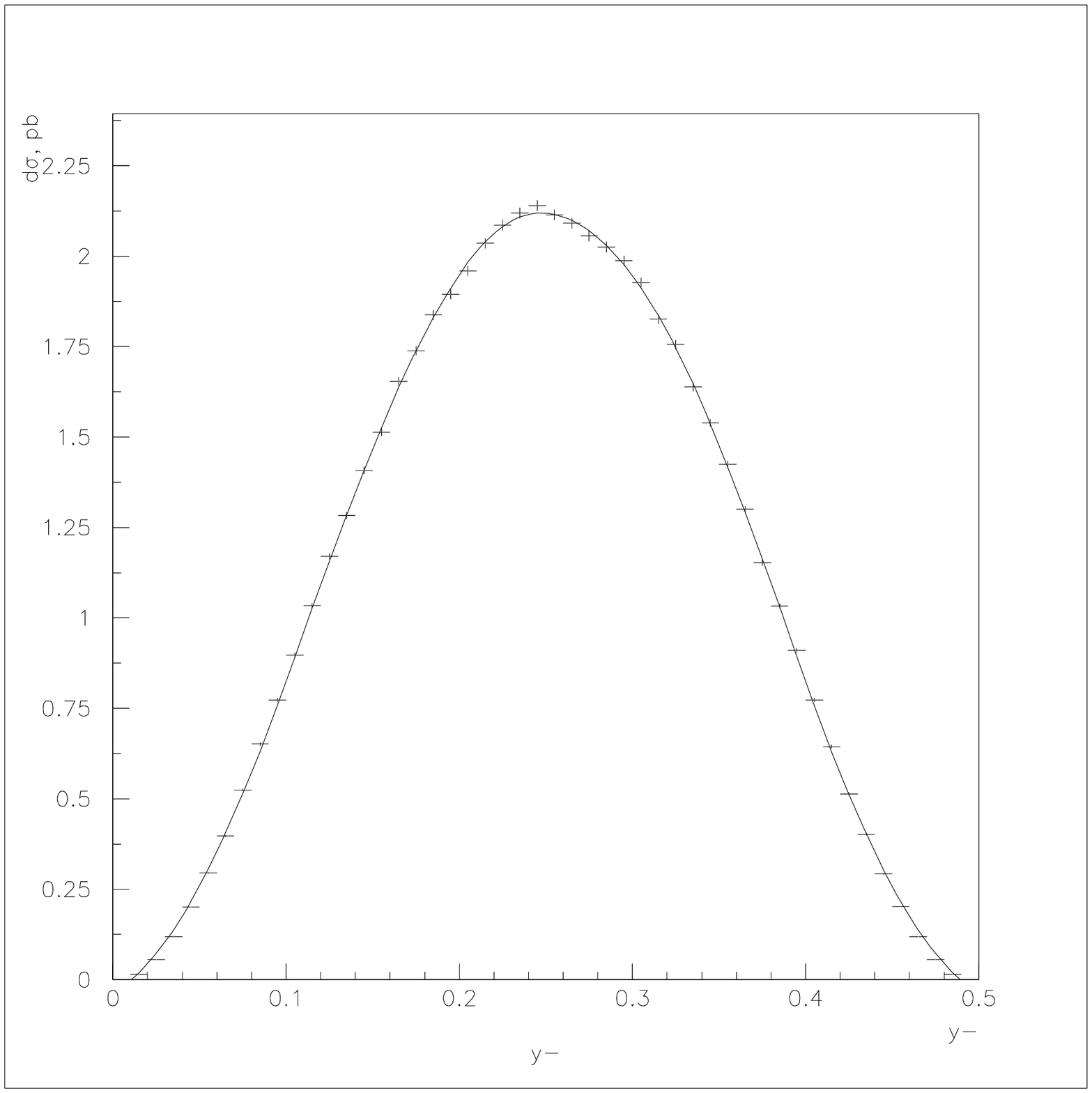}
\caption{The differential cross section $d\sigma_{+,-,0,0}/dy_-$}\label{p17}
\end{minipage}
\end{figure}

\newpage 
\begin{center}
\begin{figure}[h!]
\begin{minipage}[b]{1.\linewidth}
\centering
\includegraphics[width=0.5\linewidth, height=3.8in, angle=0]{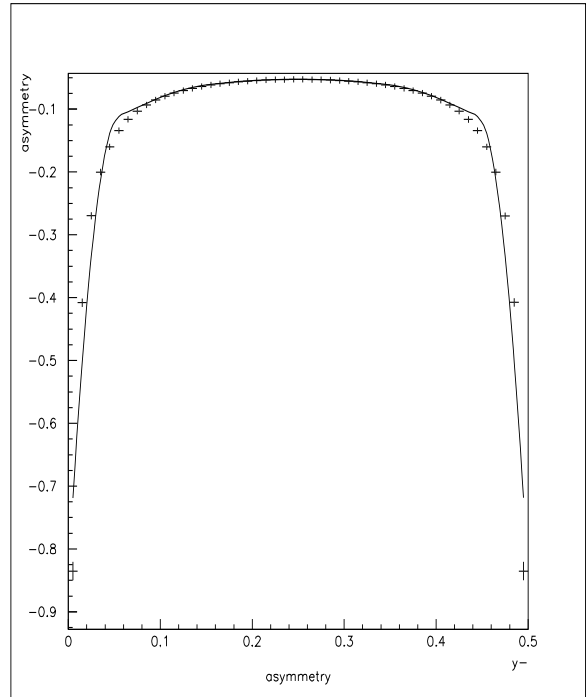}
\caption{The asymmetry $A_3$}\label{p40}
\end{minipage}
\end{figure}
\end{center}
\begin{figure}[h!]
 \leavevmode
\begin{minipage}[b]{.475\linewidth}
\centering
\includegraphics[width=\linewidth, height=3.8in, angle=0]{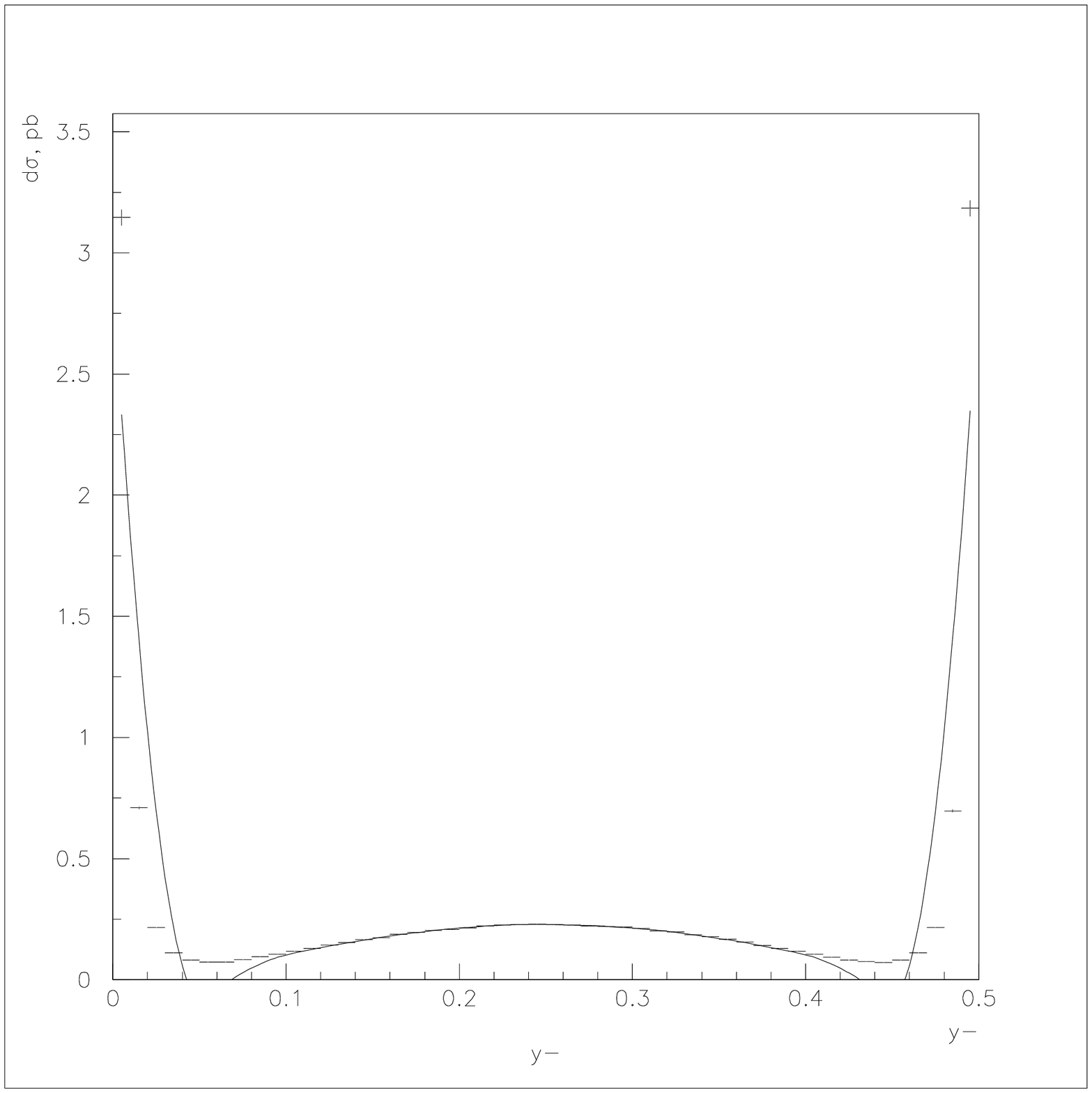}
\caption{The differential cross section $d\sigma_{+,+,unpol}/dy_-$}\label{p41}
\end{minipage}\hfill
\begin{minipage}[b]{.475\linewidth}
\centering
\includegraphics[width=\linewidth, height=3.8in, angle=0]{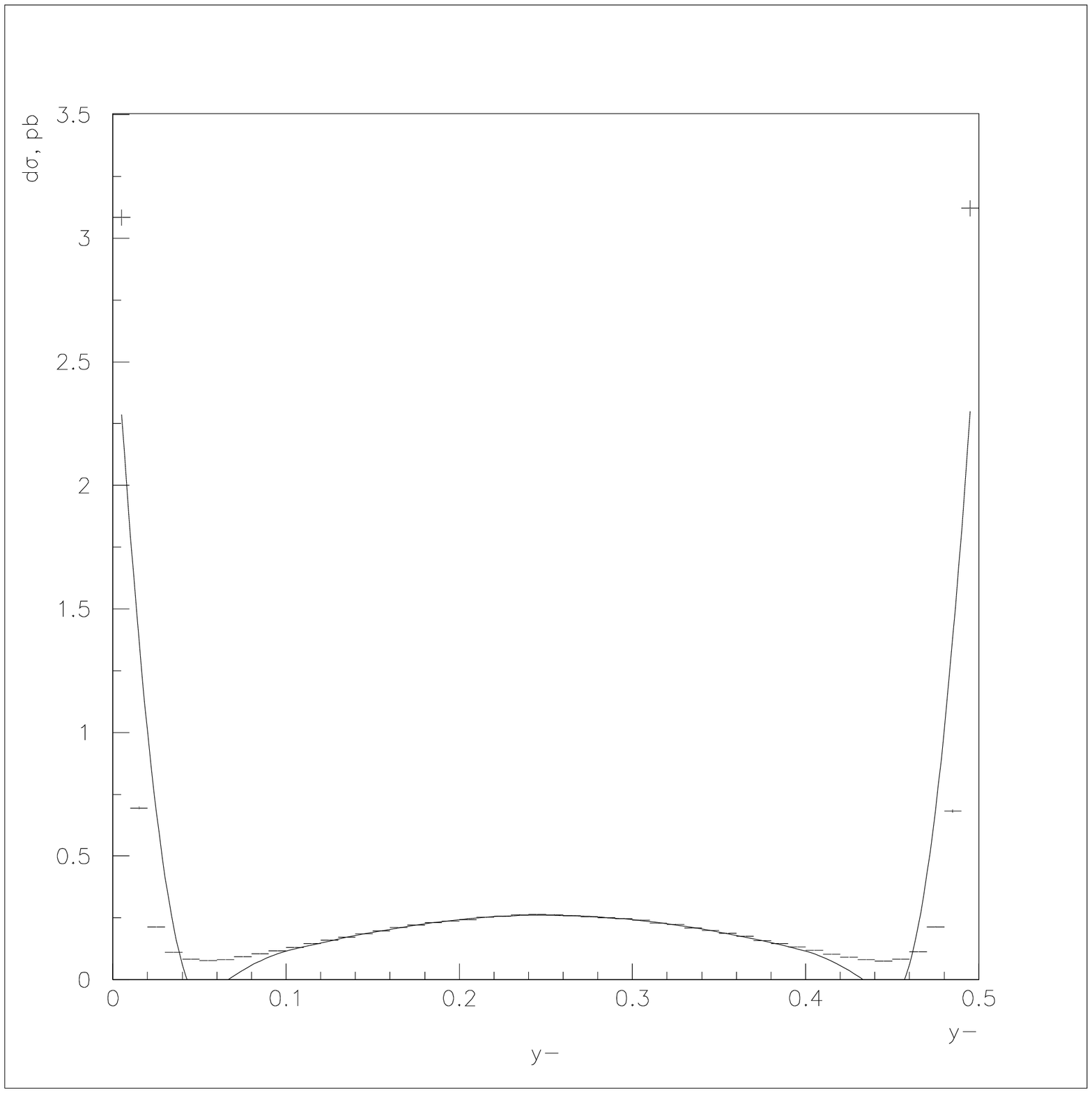}
\caption{The differential cross section $d\sigma_{+,-,unpol}/dy_-$}\label{p42}
\end{minipage}
\end{figure}
\newpage 
\begin{center}
\begin{figure}[h!]
\begin{minipage}[b]{1.\linewidth}
\centering
\includegraphics[width=0.5\linewidth, height=3.8in, angle=0]{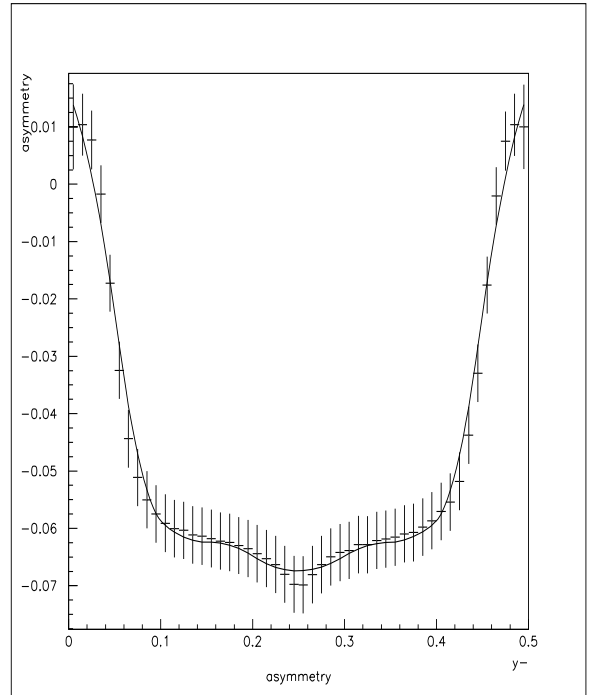}
\caption{The asymmetry $A_4$.}\label{p42}
\end{minipage}
\end{figure}
\end{center}

\begin{figure}[h!]
 \leavevmode
\begin{minipage}[b]{.475\linewidth}
\centering
\includegraphics[width=\linewidth, height=3.8in, angle=0]{+++-.eps}
\caption{The differential cross section in Born approximation $d\sigma{+,+,+,-}/dy-$.}\label{p13}
\end{minipage}
\begin{minipage}[b]{.475\linewidth}
\centering
\includegraphics[width=\linewidth, height=3.8in, angle=0]{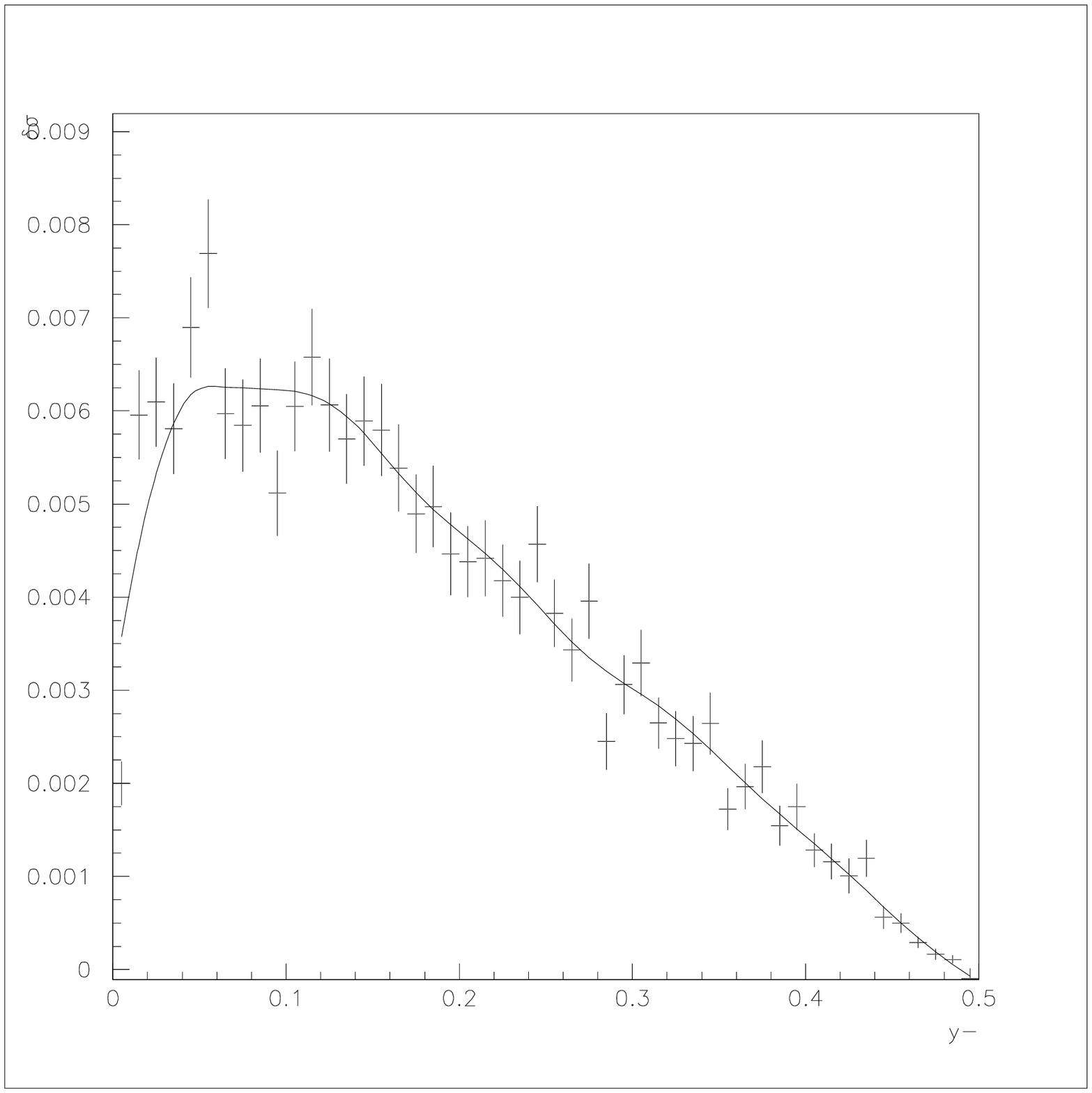}
\caption{The radiative correction for
$d\sigma{+,+,+,-}/dy-$.}\label{p12y}
\end{minipage}\hfill
\end{figure}
\newpage 
\begin{figure}[h!]
 \leavevmode
\begin{minipage}[b]{.475\linewidth}
\centering
\includegraphics[width=\linewidth, height=3.8in, angle=0]{++00.eps}
\caption{The differential cross section in Born approximation $d\sigma{+,+,0,0}/dy-$.}\label{p17}
\end{minipage}
\begin{minipage}[b]{.475\linewidth}
\centering
\includegraphics[width=\linewidth, height=3.8in, angle=0]{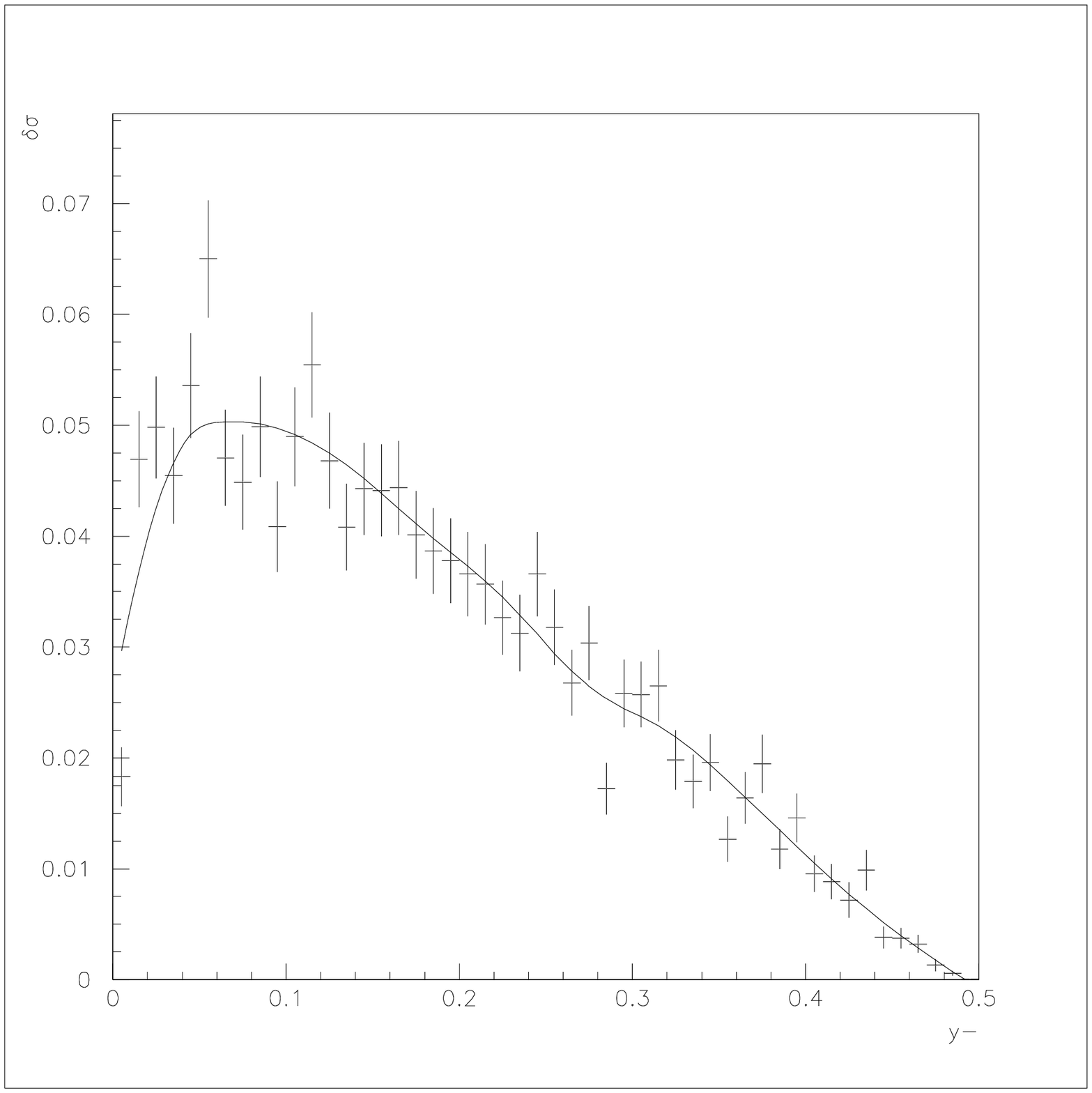}
\caption{The radiative correction for
$d\sigma{+,+,0,0}/dy-$.}\label{p16}
\end{minipage}\hfill
\end{figure}


\vspace{5mm}

\begin{thebibliography}{100}

\bibitem{bib0}
I. Marfin, V. Mossolov, T. Shishkina hep-ph/0304250  
\bibitem{bib3}
TESLA Technical Design Report Part 6. Chapter 1. Photon
collider at TESLA, hep-ex/0108012.
\bibitem{bib1}
F. Brandt  et al., Nucl. Phys.  B450 (1994) p.223. hep-ph/9308353.
\bibitem{bib2}
S. Weinzierl, NIKHEF-00-012. hep-ph/0006269.
\bibitem{bib01}
M.Shifmanet al. Sov J. Nucl. Phys. 30(1979) p.711.
\bibitem{bib02}
E.Boos et al. Phys. Lett. B275(1992) p.164.
\bibitem{bib03}
H.Veltman, Z.Phys. C62(1994) p.235.
\bibitem{bib4}
M. Roth, A. Denner Nucl. Phys.  B479 (1996) p.495. hep-ph/9605420.
\bibitem{bib5}
A. Denner, S. Pozzorini Eur. Phys. J.  C18 (2001) p.461. hep-ph/0010201
\bibitem{bib6}
A. Denner, S. Pozzorini Eur. Phys. J.  C21 (2001) p.63. hep-ph/0104127.
\bibitem{bib7}
M. Melles  Phys. Rept. 375 (2003) p.219. hep-ph/0104232.
\bibitem{bib8}
F. Jegerlehner, M. Kalmykov, O. Veretin  Nucl. Phys. B641 (2002) p.285. hep-ph/0105304.
\bibitem{bib9}
B. Kniehl Int. J. Mod. Phys. A10 (1995) p.443. hep-ph/9410330.


\end{thebibliography}
\end{document}